\documentclass[a4paper,11pt]{article}
\usepackage{pos}
\usepackage{caption}
\usepackage{subfigure}
\usepackage{natbib}
\title{Magnetohydrodynamic turbulence and propagation of cosmic rays: theory confronted with observations}
 \ShortTitle{Turbulence \& Cosmic ray propagation}

\author*[a, b]{Huirong Yan}

\affiliation[a]{Deutsches Elektronen-Synchrotron DESY, \\
Platanenalle 6, Zeuthen, Germany}

\affiliation[b]{Institute fur Physik und Astronomie, Universitat Potsdam,\\
Karl-Liebknecht-Str 24/25, Potsdam, Germany}

\emailAdd{huirong.yan@desy.de}

\abstract{Cosmic ray propagation is determined by the properties of interstellar turbulence. The multiphase nature of interstellar medium (ISM) and diversity of driving mechanisms give rise to spatial variation of turbulence properties. Meanwhile, precision astroparticle experiments pose challenges to the conventional picture of homogeneous and isotropic transport of cosmic rays (CRs). We are opening a new horizon for CR propagation research when studies of particle transport and interstellar turbulence confront each other. Here we review our recent developement on understandings of magnetohydrodynamic (MHD) turbulence and its connection to the fundamental processes governing cosmic ray propagation, different regimes of particle transport, that are augmented with observational discovery and analysis from multi-wavelength observations.}

\FullConference{37$^{\rm{th}}$ International Cosmic Ray Conference (ICRC 2021)\\
		July 12th -- 23rd, 2021\\
		Online -- Berlin, Germany}

\begin{document}
\maketitle

\section{Introduction}

Cosmic rays are energetic particles from space. Different from their electromagnetic counterpart, the distribution of CRs’ arrival direction is remarkably isotropic, with the degree of anisotropy less than 1 part in $10^3$. The frequent scattering along their paths makes it impossible to pinpoint their origins. While some favourable objects have been proposed as the sources of CRs, e.g., supernova remnants (SNRs) and pulsar wind nebulae for Galactic CRs, and extragalactic objects such as active galactic nuclei (AGN) and gamma ray bursts (GRBs) for ultra high energy CRs (UHECRs), big questions as their origin/acceleration and their propagation remain. Being charged, the CRs’ propagation is dominated by their interactions with the magnetic perturbations in space. Most of the perturbations are in the form of plasma turbulence due to the vast size span of the astrophysical system, which indicates a large "Reynolds number", the ratio of the time required for viscous forces to slow down the flow to the eddy turnover time. 

The interstellar medium is turbulent on scales ranging from AUs to kpc \cite[see][]{Armstrong95,ElmegreenScalo,Chep2010}, with an embedded magnetic field that influences almost all of its properties. Both the spectra of CRs and the ISM turbulence show big power laws, suggesting a strong interrelation of the two (Jokipii 2001). The interaction of CRs with magnetohydrodynamic (MHD) turbulence is the accepted principal mechanism to scatter and isotropize CRs \cite[see e.g.][and ref. therein]{Ginzburg1966, Jokipii1966, Wentzel74, Schlickeiser2002, Yan15}. For ultra high energy CRs (UHECR), the identification of origin also depends on the understanding of the deflection degree caused by turbulent magnetic field during the propagation in the intergalactic medium.

In addition, efficient scattering is essential for the acceleration of CRs. The escape process is inseparable from acceleration in answering the basic question of CRs' origin. Undoubtedly, the gamma ray observation is a big leap forward compared to the integrated CR spectrum we observed at earth, which is an average over long propagation times and over numerous sources. Even so, the gamma ray observation does not give us the CR spectrum in situ at the acceleration site directly, especially in the hadronic scenario, where the gamma rays emission is a convolution of the CRs escaping from their accelerator to the target gas with the original spectrum from acceleration sites. To clarify the whole process, we need knowledge of both the target gas distribution as well as the escaping process. What determines the escaping process? Is it the pre-existing turbulence in the interstellar medium? Or is it due to self-generated perturbations? What is the corresponding diffusion coefficient? The energy dependence of the diffusion is particularly important in order to decipher the originally accelerated particle spectrum at the SNR shocks.

Dark matter can generate gamma rays through the self- annihilation of the weakly interacting massive particles (WIMPs), which is so far the only direct channel for dark matter detection. The issue of dark matter direct identification is closely linked to the propagation of cosmic rays. The main uncertainties are of astrophysical origin. Arguments in favour of dark matter annihilation have been made based on the observational results including the 511keV emission from the Galactic center by INTEGRAL as well as the positron excess as detected both by PAMELA \citep{Adriani2009Nat} and by AMS02 \citep{Aguilar2013}. At the same time, astrophysical origins also seem  feasible with modified propagation models. It is essential, therefore, to have a clear picture of in-situ CR propagation from both higher resolution observations with next generation instrument like CTA and the physically grounded modelling of the particle propagation process. 

The dynamics of cosmic rays in magnetohydrodynamic (MHD) turbulence holds the key to all high energy phenomena, ranging from solar flares to remote cosmological objects such as $\gamma$ ray bursts (GRBs) \cite[see e.g.][]{YLP08, ZY11}. Recent years have seen rapid progress in cosmic ray physics, thanks to the new generation experiments. Excellent data on secondary radiations including both $\gamma$ ray and synchrotron emissions are being collected in addition to high precision measurements of local CR spectra. Many researchers build sophisticated complex models to confront the swiftly growing data sets. This makes it urgent that we understand the key underlying physical processes, can parameterize them and, if necessary, use as a sub grid input in our computer models. 

While propagation of CRs has been best understood in terms of the diffusion of cosmic rays in turbulent magnetic field, as indicated by the high degree of isotropy and the long age revealed through the relative abundance of secondary nuclei compared to their parent primary nuclei (most notably boron versus carbon) and the abundances of unstable isotopes, the precise picture is far from clear, nevertheless. The overall energy to the 1/3 power dependence seems to be compatible with an isotropic Kolmogorov turbulence. The rough consistency between classical theory and earlier CR observations sounded satisfactory until mid-90s, when the classical picture was challenged both observationally and theoretically.

Observationally, new generation CR experiments such as PAMELA, AMS02 and CREAM have discovered various anomalies deviating from the classical scenario. Examples of the challenges include the hardening of both primary CRs R$\sim 300$GeV as \citep{Ahn2010,Adriani2011,Aguilar2015a,Aguilar2015b} and secondary CRs $R\sim 200$GeVs \citep{Aguilar2018}, etc. In fact, observations have indicated that the diffusion in the interstellar medium is neither isotropic nor homogeneous \cite[see e.g.][]{Evoli2012}, which did not exist in the conventional models, but is a natural outcome of the dominance of compressible fast modes turbulence in scattering CRs as predicted by the theoretical studies based on tested model of MHD turbulence \cite{YL04,YL08}. 

From theoretical point of view, Alfvenic turbulence was demonstrated to be highly anisotropic, making it completely ineffective for scattering particles \citep{GS95, Chandran00, YL02}. The solution rests on thorough theoretical understanding of the basic interaction processes between CRs and turbulence as well as first-hand knowledge of interstellar turbulence. Multi-wavelength observations and advanced data analysis augmented with new tools based on up-to-date theoretical understandings are crucial. 

\section{What do we know about turbulence now?}

Our view on the transport of CRs has been rapidly changing, largely thanks to the advances in MHD turbulence \cite{Schlickeiser2002, Yan15}. MHD turbulence can be decomposed and the interaction of turbulence with CRs can be studied separately in each of the three MHD modes, Alfven, fast and slow, the latter of two are compressible modes \cite{CL03, MY20}. It has been demonstrated based on the tested model of turbulence that the scattering of CRs ($\gtrsim$ 100 GeV) is dominated by fast modes instead of the often-adopted Alfven modes, which indicates the inhomogeneous scattering of CRs \cite{YL02, YL04, YL08} and inefficiency of scattering on low energy CRs. For the CRs ($\lesssim$ 100 GeV), plasma instabilities play more important role~\cite{FG04,YL04,YL11}. 

\subsection{Theoretical understanding of MHD turbulence}

Turbulence in the interstellar medium was first identified by the measurement of density fluctuations, indicating the presence of compressible turbulence~\cite{ArmstrongRickett1995}. Since astrophysical plasma turbulence is compressible with finite plasma $\beta\equiv P_{gas}/P_{mag}$, the ratio between gas pressure to magnetic pressure, the magnetosonic modes should be considered when studying such turbulence. Studies have shown that the energy spectrum and the scale-dependent anisotropy of slow modes are quite similar to Alfv{\'e}n modes~\cite{LG01, CL03, MY20}. On the other hand, fast modes seem to show an isotropic cascade. This has led to important implications for astrophysical turbulence. For instance, it has been shown that fast modes are the most effective scatterers of cosmic ray particles~\citep{YL02, YL04,YL08}. 

Particle scattering and diffusion critically depends on the nature of these MHD modes. While fast modes can play an important role in scattering of cosmic rays, simulations have shown that the fast modes might only be a marginal component of compressible turbulence. However, these simulations have been driven incompressively by solenoidal forcing~\citep{VestutoOstriker2003, CL03, YangZhang2018, KowalLazarian2007}. It is important to identify whether and how changing the nature of the forcing affects the mode composition of turbulence. The question can be tackled by driving turbulence with generalized forcing, which can be always be decomposed in to solenoidal and compressive forcing. Earlier work have studied driven turbulence with a mixture of solenoidal and compressive velocity field at large scales~\citep{YangShi2016} or by decomposing the driving force into solenoidal and compressive components~\citep{FederrathRomanDuval2010}. A similar forcing in \citep{MY20} is adopted with focus on the MHD mode decomposition. The setup is highly relevant for astrophysical plasmas. It is found that the type of driving plays an essential role in determining the modes composition of turbulence.

\begin{table}[h]
\caption{\label{tab:table1}%
Simulation parameters in steady state with simulation IDs. The energy injection rate $E_{inj}$, the plasma $\beta$, Alfv{\'e}n Mach number $M_A\equiv \delta V/v_A$, sonic Mach number $M_S\equiv \delta V/c_S$, the forcing correlation time $T$, resolution, and fraction of compressive driving $\zeta$ is varied amongst the different simulation runs. From \cite{MY20}.}
\begin{tabular}{cccccccc}
\hline
{ID} & {$E_{\text{inj}}$} & {$\beta$} & {$M_A$} & {$M_S$} & T & {Resolution} & {$\zeta$}\\
\hline
S1a & $10^{-8}$ & 2.17 & 0.24 & {0.23} & 20 &$512^3$ & 1.0 \\
S2a & $8\times 10^{-8}$ & 2.17 & 0.46 & {0.44} & 10 & $512^3$ & 1.0 \\
S3a & $5\times 10^{-7}$ & 2.17 & 0.69 & {0.66} & 7.5 & $512^3$ & 1.0 \\
S4a & $8\times 10^{-6}$ & 2.17 & 0.99 & {0.95} & 5 & $512^3$ & 1.0 \\
C1a & $3\times 10^{-7}$ & 2.17 & 0.22 & {0.21} & 20 & $512^3$ & 0.1 \\
C2a & $5\times 10^{-6}$ & 2.17 & 0.48 & {0.46} & 10 & $512^3$ & 0.1 \\
C4a & $9\times 10^{-5}$ & 2.17 & 1.03 & {0.99} & 5 & $512^3$ & 0.1 \\
CB0a & $8\times 10^{-6}$ & 0.5 & 0.51 & {1.02} & 10 & $512^3$ & 0.1 \\
CB1a & $3\times 10^{-6}$ & 8.0 & 0.60 & {0.3} & 10 & $512^3$ & 0.1 \\
S1b & $10^{-8}$ & 2.17 & 0.25 & {0.24} & 20 &$1024^3$ & 1.0 \\
S2b & $8\times 10^{-8}$ & 2.17 & 0.48 & {0.46} & 10 & $1024^3$ & 1.0 \\
S3b & $5\times 10^{-7}$ & 2.17 & 0.72 & {0.69} & 7.5 & $1024^3$ & 1.0 \\
S4b & $8\times 10^{-6}$ & 2.17 & 0.87 & {0.84} & 5 & $1024^3$ & 1.0 \\
C1b & $3\times 10^{-7}$ & 2.17 & 0.23 & {0.22} & 20 & $1024^3$ & 0.1 \\
C4b & $9\times 10^{-5}$ & 2.17 & 1.05 & {1.01} & 5 & $1024^3$ & 0.1 \\
\hline
\end{tabular}
\end{table}

\begin{figure}[h]
\centering 
\begin{subfigure}[]{}
         \centering
         \includegraphics[width=0.48\textwidth]{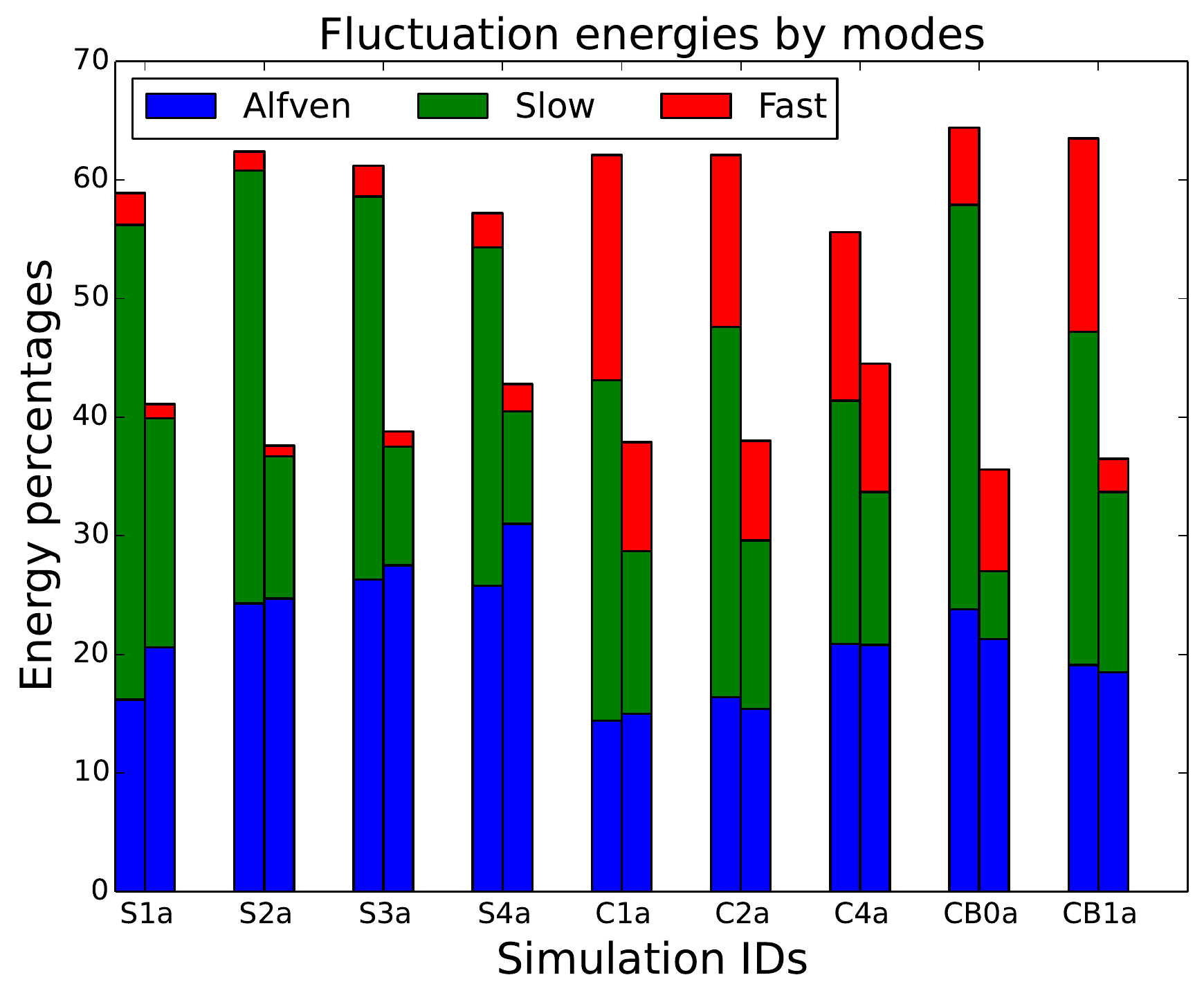}
              \end{subfigure}
     \hfill
\begin{subfigure}[]{}
         \centering
         \includegraphics[width=0.48\textwidth]{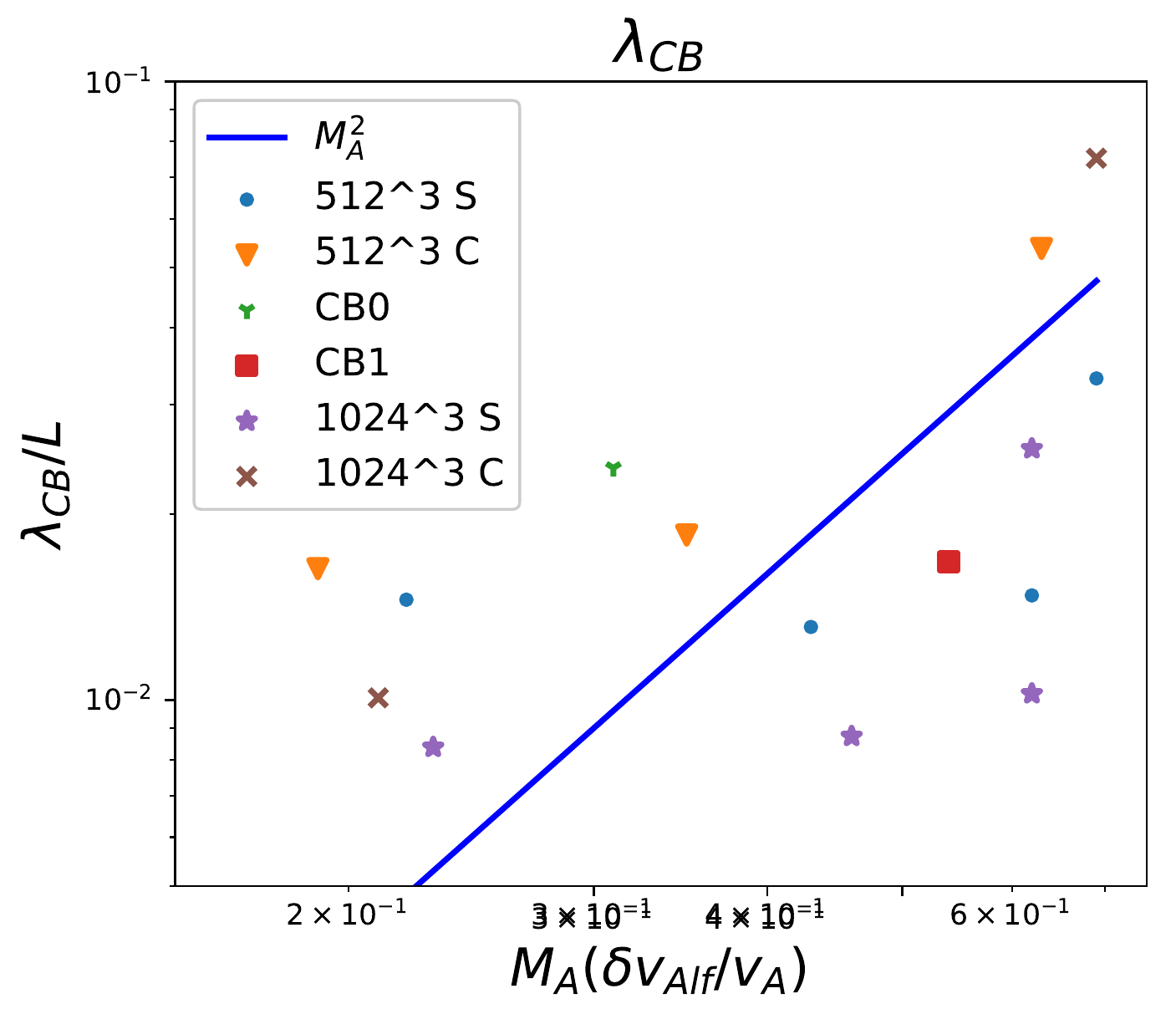}
              \end{subfigure}
\caption{{\it a)} The time-averaged fractions of mode energies in different modes in different simulations. Each simulation has two bars, the left one represents the velocity field showing the three mode percentages ($P_{KEA}$, $P_{KES}$, and $P_{KEF}$ in blue, green, and red respectively). Similarly the right hand bar is for the magnetic field showing $P_{MEA}$, $P_{MES}$, and $P_{MEF}$ in their respective colors. Both the bars add up to 100\%. Compressive driving leads to a significantly larger fraction of the fast magnetosonic mode. {\it b)} The variation of the estimated transition scale of weak to strong turbulence $\lambda_{CB}$ with the Alfv{\'e}nic Mach number $M_A$. The dots are the results from the different simulations. The blue line is showing the $M_A^2$ reference line. From \cite{MY20}.}
\label{energy_fracs_barplot}
\end{figure}

The mode energy fractions are shown in Fig.~\ref{energy_fracs_barplot} with information of each turbulence simulation listed in Table\ref{tab:table1}. In the solenoidally driven simulations, the Alfv{\'e}n and slow modes form the major fraction, with very little contribution from fast modes ($\sim 5\%$ in the tran-Alfv\'enic case). The Alfv{\'e}n modes have roughly equal energies in the velocity and magnetic fields while the slow mode has a stronger component of velocity fields. In the compressively driven simulations, however, fast mode has a significantly large proportion, which has not been observed before (see Fig.~\ref{energy_fracs_barplot}). On the other hand, increasing the $M_A$ is not affecting the mode fractions significantly, except for a gradual increase in Alfv{\'e}n mode proportion. Kinetic fluctuations of slow modes decrease while their magnetic fraction increases as the plasma $\beta$ increases, bringing the slow mode magnetic and kinetic fluctuations closer to equipartition. This is understandable from the fact that the slow mode become pseudo-Alfv{\'e}n mode as $\beta\rightarrow\infty$. As $\beta$ increases, the fraction of fast mode rises in the kinetic fluctuations, while decreasing in the magnetic fluctuations. This is expected since fast modes are essentially sound waves in the high $\beta\rightarrow\infty$ limit. In all the compressively driven simulations, the total energy in slow and fast modes is larger than that of Alfv{\'e}n mode. 

Understandably, compressive driving leads to a larger proportion of the slow plus fast magnetosonic modes, but more specifically the fast magnetosonic modes. This is crucial for the cosmic ray transport in turbulence since fast modes dominate the particle scattering~\citep{YL02,YL04}.

\begin{figure}[h]
\centering 
\includegraphics[width=1.0\linewidth]{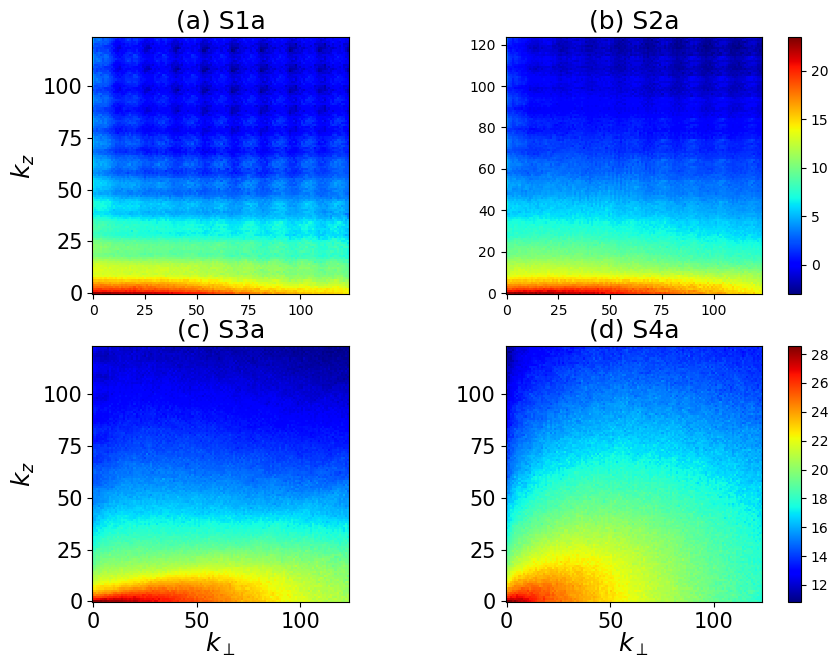}
\caption{The $k_{z}-k_{\perp}$ wavenumber spectrum for the velocity field of Alfv{\'e}n modes with increasing Mach number b) fast modes. The color indicates logarithm of the spectrum power. S1a is sub-Alfv\'enic with the $M_A\sim 0.24$, and S4a is trans-Alfv\'enic. The power spreads more in the parallel direction as Alfv{\'e}n Mach number increases. From \cite{MY20}.}
\label{kspec_par_perp_A_S1-4}
\end{figure}

\begin{figure}[h]
\centering 
\includegraphics[width=1.0\linewidth]{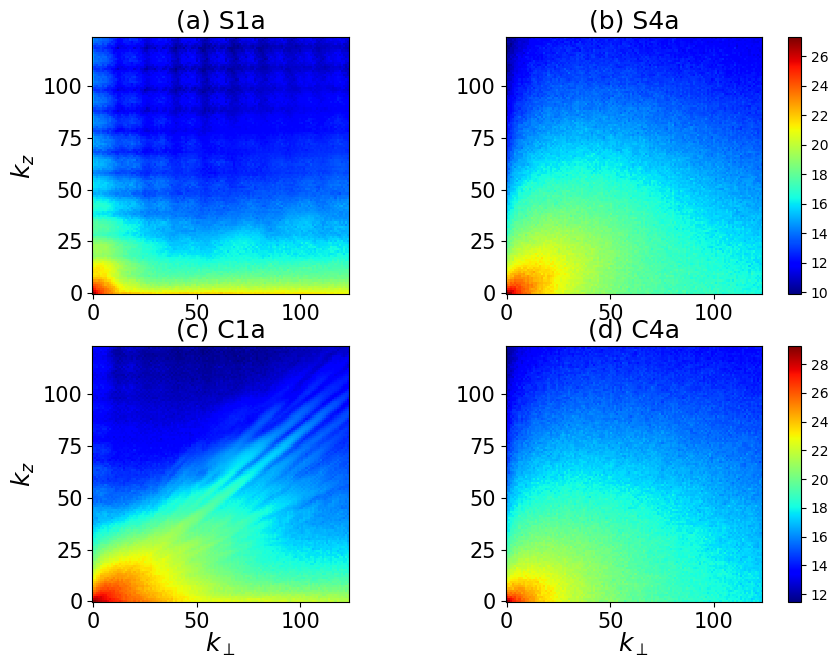}
\caption{The $k_{z}-k_{\perp}$ wavenumber spectrum for the velocity field of fast modes. The color indicates logarithm of the spectrum power. C1a is sub-Alfv\'enic with the $M_A\sim 0.22$, and C4a is trans-Alfv\'enic. Fast modes cascade show little change with $M_A$ as well as forcing scheme. From \cite{MY20}.}
\label{2Dkspec_F}
\end{figure}

Another regime less explored is low $M_A$/sub-Alfv\'enic turbulence. Theoretically, the Alfv{\'e}nic turbulent cascade is expected to be weak before transitioning on smaller scales to strong turbulence characterized by the critical balance condition~\citep{GS95}. For decaying turbulence, the transition has been studied ~\citep{MeyrandGaltier2016}. For driven turbulence, the $M_A$ dependence of this transition is explored in \cite{MY20} and it is found that the nature of Alfv{\'e}nic turbulence in the low $M_A$ regime depends also on the driving. The conventional forcing with  delta-correlation in time produces faster dynamics and therefore is unable to capture the weak cascade. Indeed with Ornstein-Uhlenbeck type of forcing, weak regimes of Alfv{\'e}nic turbulence are observed to extend from the turbulence injection scale $L_{\text{inj}}$ down to $L_{\text{inj}}M_A^2$ (Fig.\ref{energy_fracs_barplot}b), consistent with the theory. Fig.~\ref{kspec_par_perp_A_S1-4} shows the 2D spectra for velocity field of the Alfv{\'e}n modes in simulations with increasing $M_A$. It shows that for $M_A<0.5$ the energy is distributed along the $k_{\perp}$ axis close to $k_{z}=0$. There is very little cascade along the parallel direction to higher $k_{z}$. As the $M_A$ increases the cascade slowly spreads in the parallel direction, indicating the transition of the turbulence from weak to strong regime. 

A particularly intriguing question is how fast modes behave in weak turbulence. Do they also show an $M_A$ dependent behavior like Alfv{\'e}n modes in terms of weak or strong turbulence? Recent work \cite{MY20} find through performing higher resolution studies that fast modes do not exhibit any transition from the large scales to small scales and have an extended inertial range with isotropic cascade regardless of Alfv\'enic Mach number in contrast to the case of Alfv{\'e}n and slow modes. Fig.~\ref{2Dkspec_F} shows the 2D spectrum of fast mode. The spread of energy for the fast mode appears very close to isotropic. The isotropic nature of the fast mode cascade is similar with both solenoidal and compressive driving. This shows that the isotropic nature of fast mode cascade is a robust feature. 

As the result, cosmic ray scattering and acceleration remain effective in sub-Alfv\'enic turbulence, particularly in the regime dominated by compressible driving, through both the gyroresonance and transit-time damping interactions with fast modes. The results of \cite{MY20} also suggest that these different mode cascades are not completely independent of each other, depending on which mode is dominant. The spectrum of the fast modes can be steeper than $k^{-3/2}$ and closer to $k^{-2}$ when the fast mode dominates. This has implications for the cutoff scale and damping of fast modes. The nature of turbulence can be different depending on local driving and environment and this has important implications on related problems. 

\subsection{Observational advances on MHD turbulence}
\label{turb_obs}
\begin{figure}
\includegraphics[width=\columnwidth]{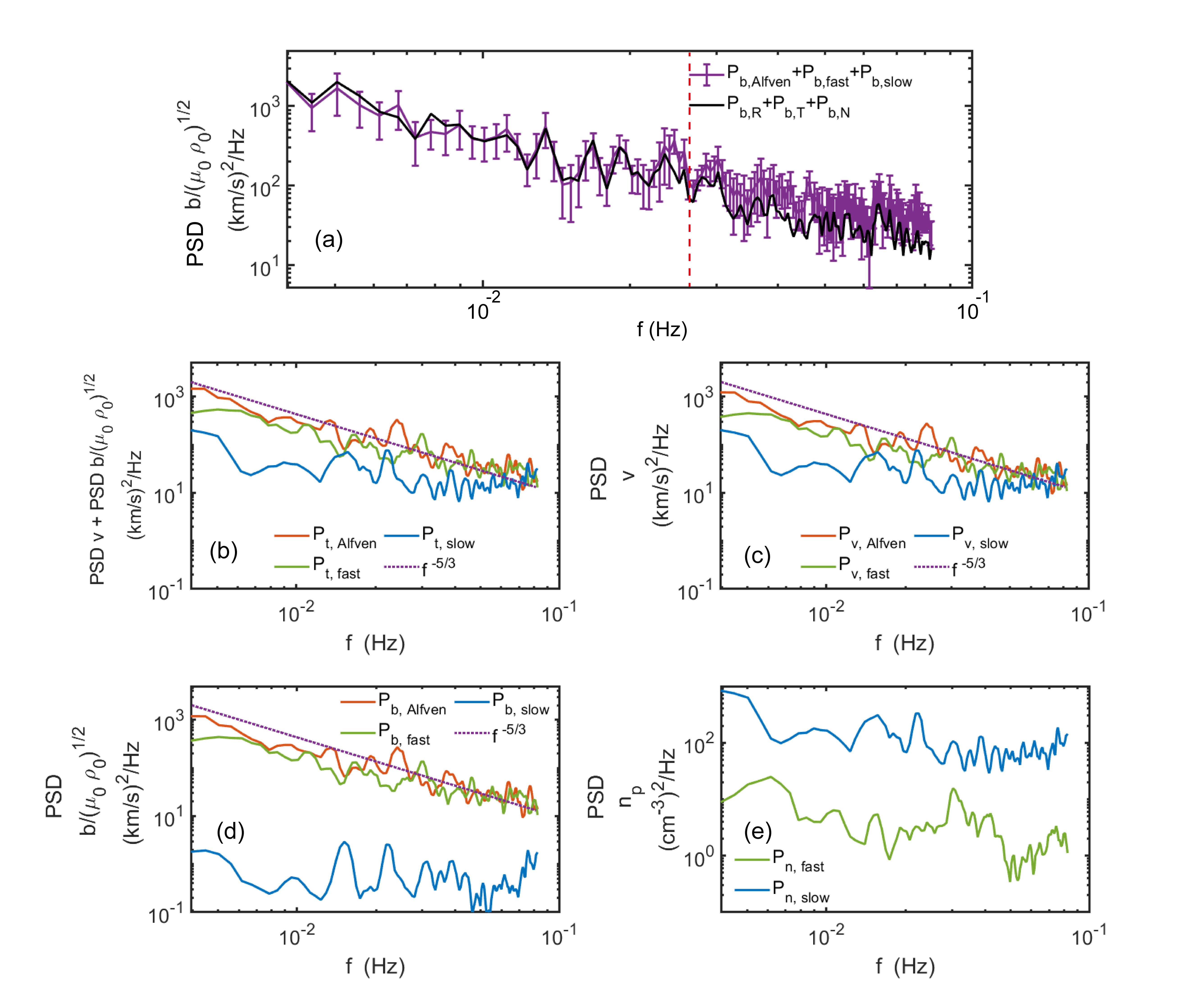}
\caption{Power spectral densities from three MHD modes in the spacecraft frame during 19:30:30-20:14:30 UT. {\em a)} the sum of magnetic power: the observed magnetic power (black; $P_{b,obs}$ and the magnetic power calculated based on the ideal MHD theory (purple; $P_{b,3modes}=P_{b,Alfven}+P_{b,fast}+P_{b,slow}$. The red vertical dashed line represents the frequency $f_{sc}~0.026$ Hz. The error bars stand for the standard deviation. {\em b)} total power spectra; {\em c)} kinetic power spectra; {\em d} magnetic power spectra; {\em e} proton density power spectra. Red, green, and blue curves represent Alfvén, fast, and slow modes, respectively. The purple dashed lines in (b-d) mark the Kolmogorov-like power law ($f_{sc}^{-5/3}$) as a reference. From \cite{Zhao2021}.}
\label{PSP}
\end{figure}

It is absolutely essential to determine basic parameters of interstellar turbulence from observations, determined by local environmental conditions. In space plasma, extensive studies on solar wind turbulence have been conducted based on in situ measurements. For instance, detailed analysis of Parker Space Probe (PSP) data have disclosed the existence of both Alfv\'en and compressible modes \cite{slowmode_shi15, Chaston2020, Zhao2021}. In particular, recent results demonstrate the quantitative energy proportion among the three MHD modes. Fast modes, although subdominant, are not negligible even in the radial solar wind known to be prevailed by Alfv\'enic fluctuations (see Fig.\ref{PSP}). With interstellar turbulence, it is much more nontrivial understandably. There are two possible sources of information about interstellar turbulence. The parameters of velocity field can be retrieved from spectrometric observations of emission lines. Radio-frequency observations of synchrotron emission (intensity and polarization) can offer valuable information about turbulent magnetic field.

The equi-partition of magnetic and thermal energy in the interstellar medium (ISM) indicates the magneto-hydrodynamic nature of the interstellar turbulence, which results in three distinctive plasma modes: Alfv\'en, fast and slow magnetosonic modes. It is inadequate to focus on the one dimensional spectrum in the inertial range in the case of MHD turbulence, which has three dimensional structures and anisotropies. There are a variety of the drivers for turbulence in ISM ranging from supernovae explosions \cite{MO77}{}, accretion flows \cite{KH10_accretion}{}, magneto-rotational instability in the galactic disk \cite{Balbus99}{}, thermal instability \cite{KN02_thermal}{}, to collimated outflows \cite{NL07_outflow}{}, etc. The diversity of driving mechanisms and multiphase nature of ISM naturally give rise to spatial variations of turbulence properties, in particular, the relative proportion of the three modes, implying the spatial inhomogeneity of CR transport. Nevertheless, the employed model of interstellar turbulence is often oversimplified, being assumed to be only Alfv\'enic or even hydrodynamic due to a lack of observational evidence.  One argument is that magnetosonic (henceforth MS) modes are subjected to severe damping. However, quantitative studies have demonstrated that the impact of damping is limited. The cascade of fast modes, for instance, can survive below sub-parsec scales in fully ionized plasma~\cite{YL04, YL08}.   

It is challenging, however, to remotely diagnose the magneotosonic modes in interstellar turbulence. A breakthrough is made recently in the observational study of the plasma modes. A technique based on polarized synchrotron emission (Synchrotron Polarization Analysis, hereafter SPA) is developed. The diffuse synchrotron radiation is generated while relativistic electrons are traveling in the magnetic field. The synchrotron radiations, particularly the polarization signals, thus carry the information of MHD turbulence since the interstellar magnetic field is turbulent. The SPA method can be used to link the synchrotron polarization properties and the underlying turbulence statistics associated with different plasma modes. The SPA method is then applied to the synchrotron polarization data from two different regions with prominent synchrotron radiation. 

\begin{figure*}
\centering
\begin{subfigure}[]{}
\includegraphics[width=0.45\columnwidth, height=0.35\textheight]{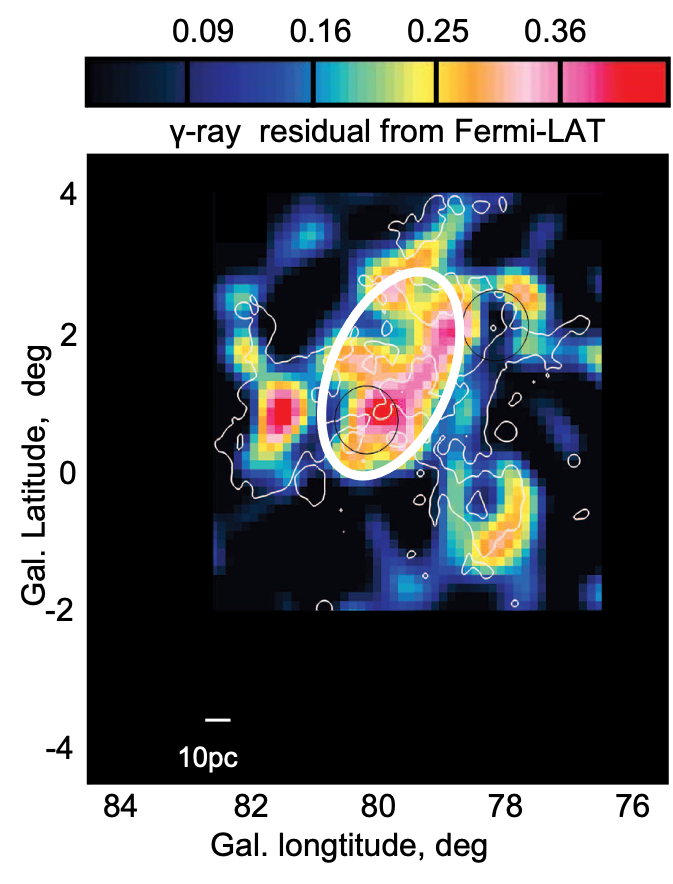}
\end{subfigure}
\begin{subfigure}[]{}
\includegraphics[width=0.45\columnwidth, height=0.35\textheight]{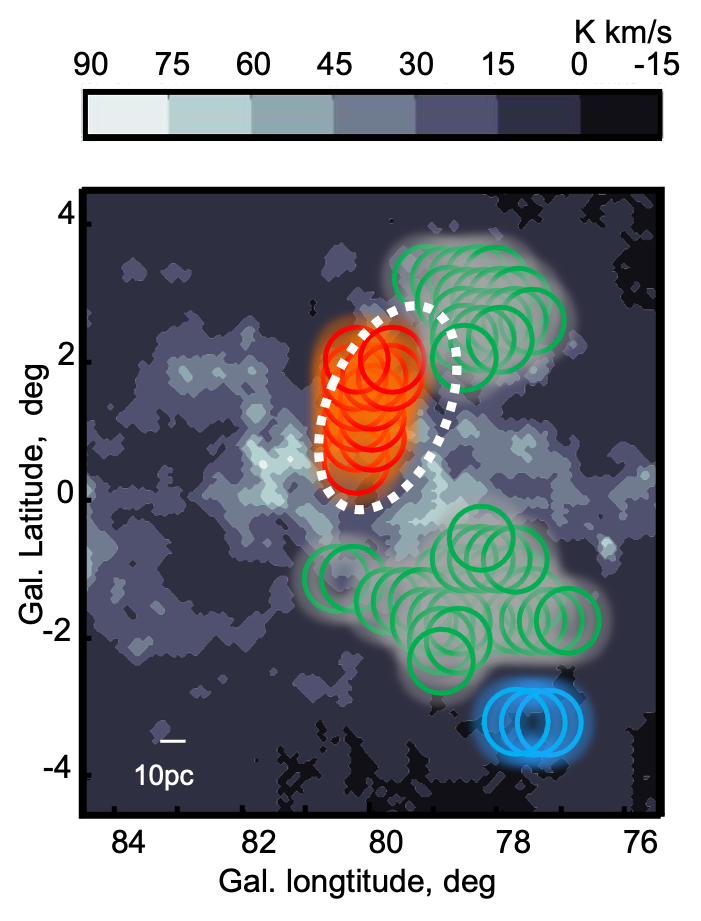}
\end{subfigure}
\\
\begin{subfigure}[]{}
\includegraphics[width=0.45\columnwidth, height=0.35\textheight]{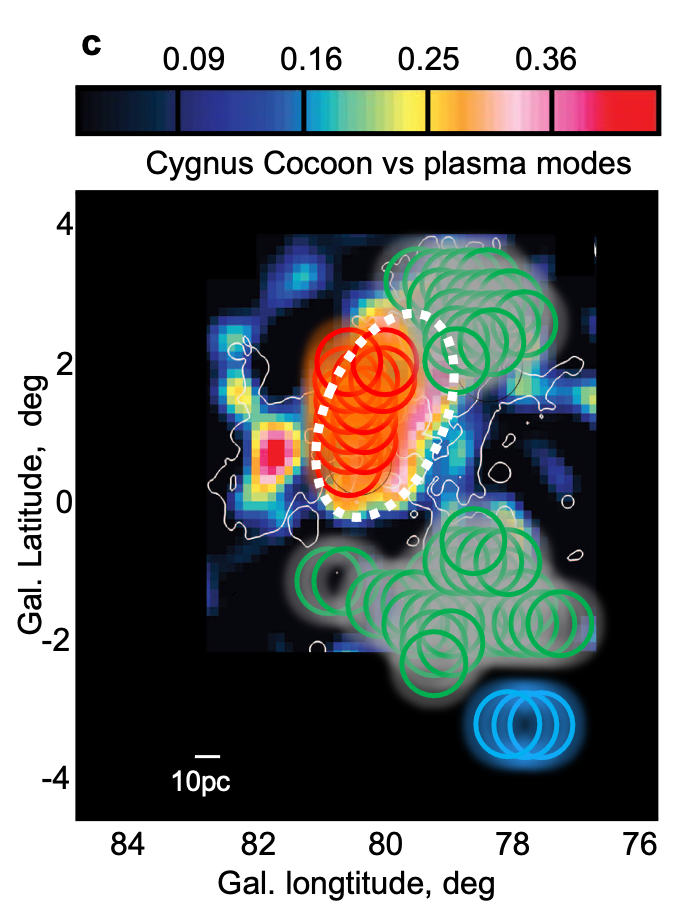}
\end{subfigure}
\hfil
\begin{subfigure}[]{}
\includegraphics[width=0.48\columnwidth, height=0.3\textheight]{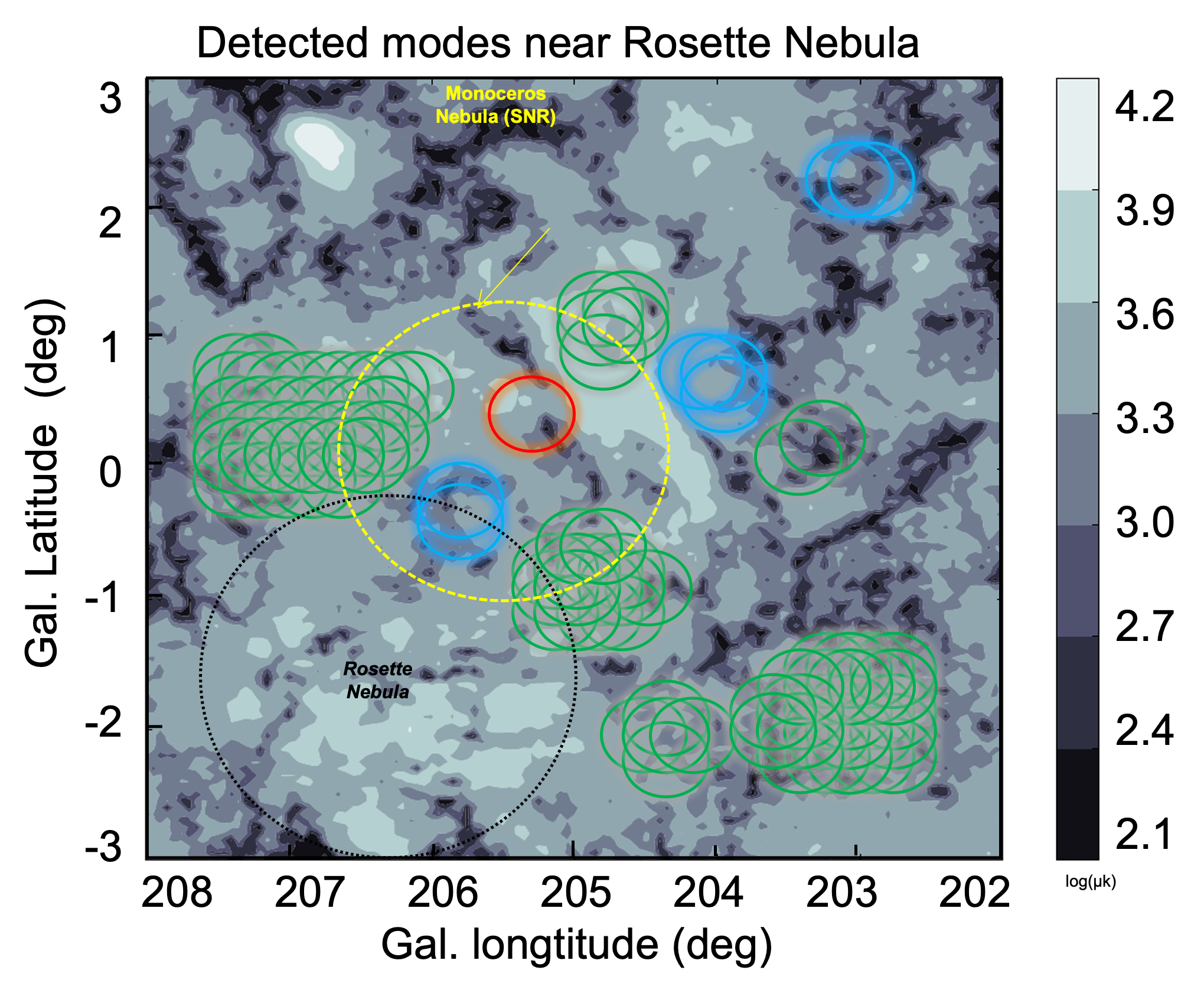}
\end{subfigure}
\caption{{ {\em a-c}) Comparison between the plasma modes identified in Cygnus X region and multimessengers.} {\em a}) Gamma ray map of Cygnus \citep{FermiLAT:2011}. {\em b}) Turbulence modes identified over the CO map of Cygnus from \citep{Dame01}. {\em c}) Turbulence modes identified in Cygnus X region plotted over the gamma ray map. The magnetosonic modes overlaps in a high consistency with extended excess of hard emission observed by Fermi-LAT, the "Fermi cocoon". {\em d)} The detected signatures compared with synchrotron intensity in the vicinity of Rosette nebula. The center of the map is SNR G205.5$+$0.5. Bottom left is Rosette Nebula. From {\cite{ZCY20}}.}
\label{Cyg}
\end{figure*}

As the result, the dominant plasma modes have been discovered for the first time in interstellar turbulence {\citep{ZCY20}}. As shown in Fig.\ref{Cyg}, different plasma modes are revealed in various Galactic medium, rendering a direct proof that interstellar turbulence is magnetized. The results of the modes identification with SPA are displayed in Fig.\ref{Cyg} for Cygnus X region, a complex of giant molecular clouds hosting massive star-forming activities with rich collection of young massive stars and supernovae \cite{Beerer10,Herschell16,Wright12}{}. The identified signatures are also overlaid on the Extinction map (see Fig.\ref{Cyg}c). The overall detected plasma modes overlap to a large extent with the important active star forming regions, Cygnus X South\cite{Schneider2011,Rygl12,Maia16}{}. The middle 2-degree zone exhibits substantial amount of MS modes. Alfv\'en modes are also discovered in the north and south regions.

Moreover, this region has also a diffuse Fermi superbubble of $\gamma-$ray excess above 3GeV (Fig.\ref{Cyg}a)~\cite{FermiLAT:2011}{}, which can be explained neither by neighboring pulsar wind nebulae nor by density enhancement as indicated by CO map (Fig.\ref{Cyg}b). In addition, intense CR emission in Cygnus cocoon is also detected by HAWC~\cite{HAWCCYG}{}. Fig.\ref{Cyg}c clearly demonstrates that the cocoon is correlated with the identified magnetosonic modes to a high degree of consistency. Interstellar turbulence has a huge span ranging from $\sim 100$pcs to $\sim 10^9$cms as observed \cite{Armstrong95}{}. The Alfv\'enic turbulence and MS modes become decoupled on scales smaller than the injection scale and form separate cascade \cite{CL03}{}. Consequently, the magnetosonic perturbations discovered on the scales of a few tens parsecs indicate that the percentage of magnetosonic modes can be much higher than other regions on all smaller scales down to dissipation. Earlier studies demonstrate that the MS modes play a dominant role in CR scatterings \cite{YL02, YL04}{}, thus providing a stronger confinement for CRs. Therefore, the observation of plasma modes unveils the origin of CR concentration in the Cygnus cocoon, and provides the first observational evidence for the dominance of magnetosonic modes on the cosmic ray (CR) transport and acceleration. The plasma modes information is evidently indispensable in the studies of CR propagation and acceleration.

Different plasma modes are also identified in the vicinity of Rosette Nebula (Fig.\ref{Cyg}d). Rosette Nebula and the supernova remnants (SNR) ${\rm G205.5+1.5}$ are located at the same distance, interacting with each other \cite{Odegard1986}{}. The Alfv\'enic signatures seen at the western and northern edges of the Rosette Nebula point to the Alfv\'en modes. Furthermore, MS modes is only observed in the center of the SNR. The MS and Alfv\'en signature dominance imply their corresponding forcing mechanisms. It is plausible that the turbulence in the center of SNR is driven compressively by the supernova shock, leading to substantial amount of compressible MS modes. On the other hand, the turbulence in the edges of the molecular cloud is probably driven by the shearing motion, resulting in the dominance of incompressible Alfv\'en modes as observed. In addition, isotropic signatures emerges in between the SNR and molecular cloud, indicating the existence of super-Alfv\'enic turbulence.

These results pinpoint the necessity to account for plasma property of turbulence, which is neither hydrodynamic nor purely Alfv\'enic, but possess different characteristics, particularly 3D anisotropy \citep{LG01,CL03} depending on local physical conditions, particularly the driving process\citep{MY20}. This study unfolds a new avenue of connecting plasma physics with macro astrophysical phenomena. Moreover, the observation of plasma modes composition in the Galaxy has far reaching consequences on not only cosmic rays and star formation, but also the fundamental understanding of the driving mechanism of turbulence. A new window of opportunity arises in the multi-messenger research, which relates directly the turbulence properties exposed by radio synchrotron signals with CRs and diffuse gamma ray emissions as well as star formation activities. A highly promising research field is foreseen to unroll with ample results anticipated from the high resolution synchrotron polarization data analysis and multiple wavelength comparison, that will shed light on the role of turbulence in various physical processes.

\begin{figure}
\centering
\begin{subfigure}[]{}
         \centering
\includegraphics[width=0.31\textwidth,height=0.2\textheight]{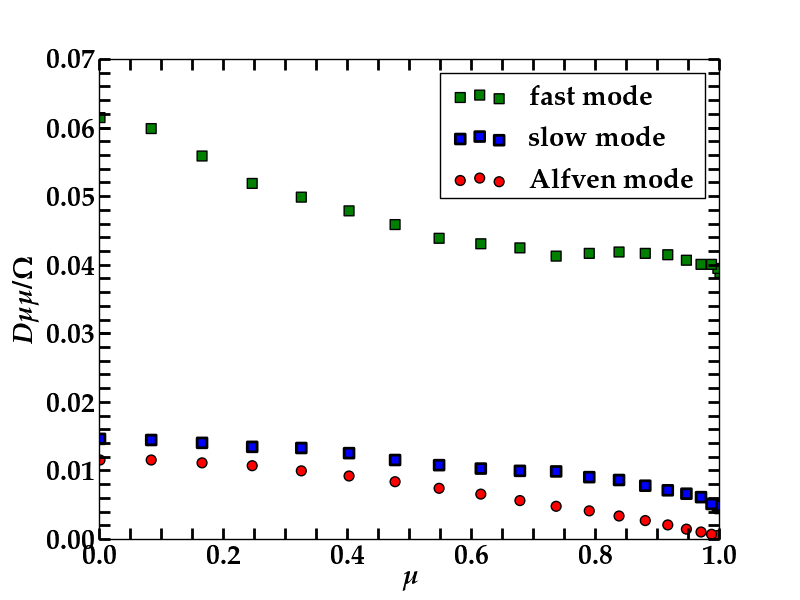}
\end{subfigure}
\begin{subfigure}[]{}
         \centering
\includegraphics[width=0.33\textwidth, height=0.2\textheight]{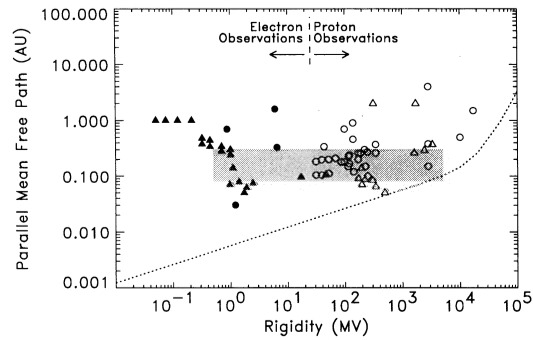}
\end{subfigure}
\begin{subfigure}[]{}
         \centering
\includegraphics[width=0.31\textwidth, height=0.2\textheight]{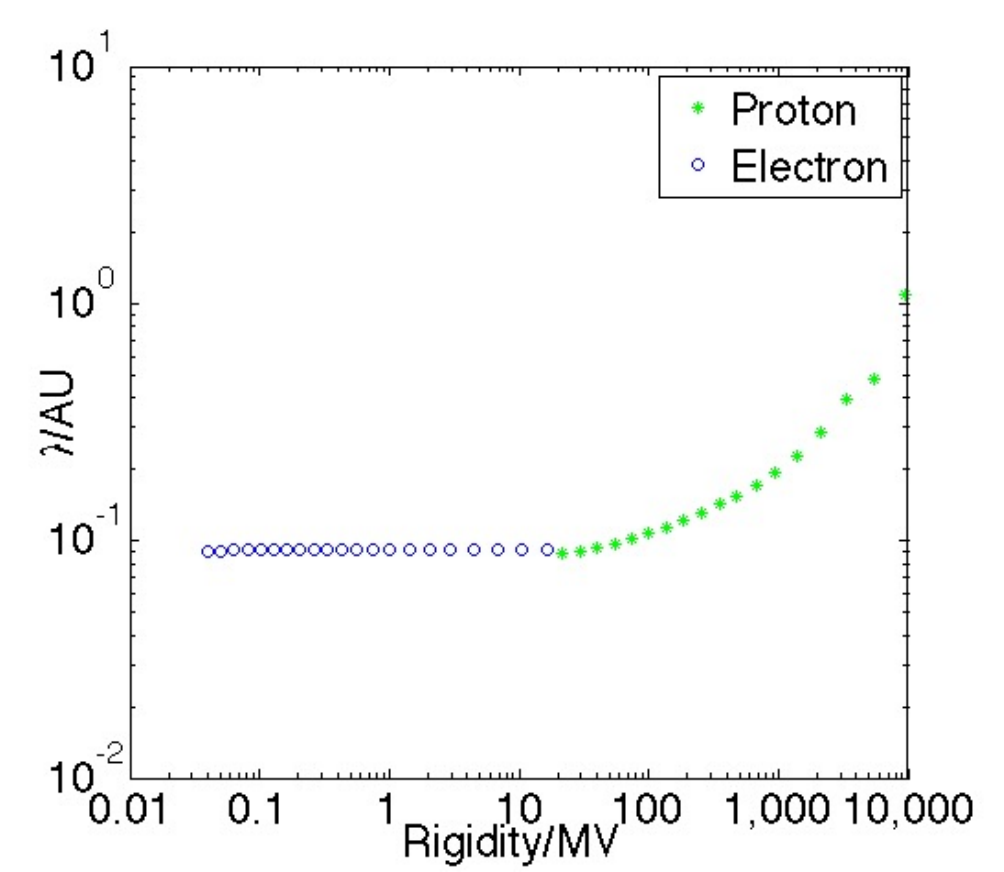}
\end{subfigure}
 \caption{{\em a)} Pitch angle diffusion coefficients for CRs in different MHD modes with $M_A\sim 0.9$. The x axis represents the initial pitch angle cosine, $\mu$. The y axis represents the pitch angle scattering coefficient normalised by the gyrofrequency, $D_{\mu\mu}/\Omega$. Different symbols represent different MHD modes: Alfv\'en (red), slow (blue) and fast (green). From \cite{Maiti2021}. {\em b)} Palmer consensus, from \cite{Bieber94}. {\em c)} the observed flat energy dependence as in {b)} is consistent with particle scattering by fast modes in collisionless medium.}
\label{SNR}
\end{figure}

\section{Particle transport: theoretical and observational studies}

\subsection{Transport properties in difference MHD modes}
\label{largescale_transport}

Cosmic ray transport is intimately linked to the property of MHD turbulence. Different from hydrodynamic turbulence, MHD turbulence is much widely diversified depending on the parameters in local interstellar environment, such as Mach number and plasma $\beta$. Another factor, that has been frequently overlooked, is the modes composition of MHD turbulence. It can vary substantially depending on the driving mechanism of turbulence \citep{MY20}. This is particularly important in view of the fact that different MHD modes contribute to CR transport differently. It is therefore inadequate to depict CR transport as that described by Kolmogorov turbulence with one characterization even in the high energy regime where external turbulence dominates the CR scattering.

The qualitative and quantitative predictions of the CR transport in MHD turbulence relies on our precise understanding of different basic contributing mechanisms from the three different plasma modes: resonant wave-particle interactions including gyro-resonance and transit time damping (TTD, mirror interaction with Landau resonance condition) and the spatial/temporal field lines separation. In the recent work \cite{Maiti2021}, the contributions of the three MHD modes are studied via test particle simulations performed in MHD turbulence data cubes. 

The pitch angle diffusion coefficients and their variation with initial pitch angle cosines is presented in Fig.\ref{SNR}a for the three MHD modes, Alfv\'en, slow and fast.
The result agrees well with the prediction of the nonlinear theory~\citep{YL08}. Compressible modes contribute to particle scattering through both gyroresonance and resonant mirror (transit time damping, TTD) interaction, the latter of which only operates with compressible modes. Alfv\'en modes, on the other and, only scatter particles through gyroresonance. This is why slow modes are slightly more efficient in scattering particles despite that they have the similar anisotropy as Alfv\'en modes. In comparison to the anisotropic Alfv\'en and slow modes, the scattering with the isotropic fast modes are more efficient. We note that the inertial range in the current MHD simulations is limited. The interstellar turbulence cascade spans more than 10 decades \citep{Armstrong95, Chep2010}. CRs experience, therefore, much more anisotropic Alfv\'enic turbulence on the resonant scales, which are 6-7 orders of magnitude smaller than the turbulence injection scale ($\sim 100pc$) in interstellar medium. This indicates the role of fast modes in scattering CRs is even more prominent in the Galactic ISM. The parallel diffusion coefficient varies, thereby, with the percentage of fast modes and the forcing mechanism of the local turbulence \cite{MY20,ZCY20}.

Note that damping plays an important role in shaping the energy dependence of diffusion. Indeed unlike the commonly advocated Alfv\'epic turbulence, compressible fast modes are subject to various damping processes which depends on the medium properties, such as plasma $\beta$. While bearing an inertial range with self-similarity, turbulence is largely shaped by forcing on large scales and damping on small scales. Damping of fast modes highly depends on the propagation angle for both collisionless damping and viscous damping~\cite{YL04}. Analytical study on the damping, shows that the effect on turbulence scattering of CRs is limited because of decreased damping on quasi-slab modes~\cite{YL02,YL08}. The difference in damping process and environmental dependence shapes the corresponding CR diffusions since gyroresonance with turbulence happens on small scales. For instance, in collisionless environment, it has been shown that the mean free path is weakly dependent on particle energy~\cite{YL08}, this provides an viable explanation to the observed flat dependence in solar wind, known as Palmer consensus~\cite{Palmer1982, Bieber94} considering that the solar wind is generally collisionless (Fig.\ref{SNR}b,c). It may also account for the harder spectrum of Cygnus cocoon by Fermi~\cite{FermiLAT:2011}, particularly in view of the dominance of MS modes identified in the region as discussed in \S\ref{turb_obs} (Fig.\ref{Cyg}c). 

\begin{figure}
\centering
  \begin{subfigure}[]{}
\includegraphics[width=0.31\textwidth,height=0.2\textheight]{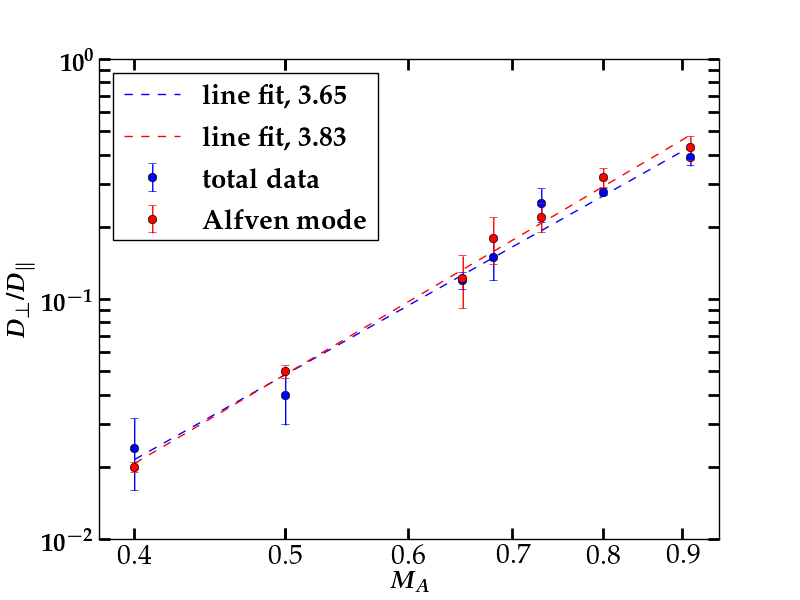}
 \end{subfigure}
 \begin{subfigure}[]{}
\centering
\includegraphics[width=0.31\textwidth,height=0.2\textheight]{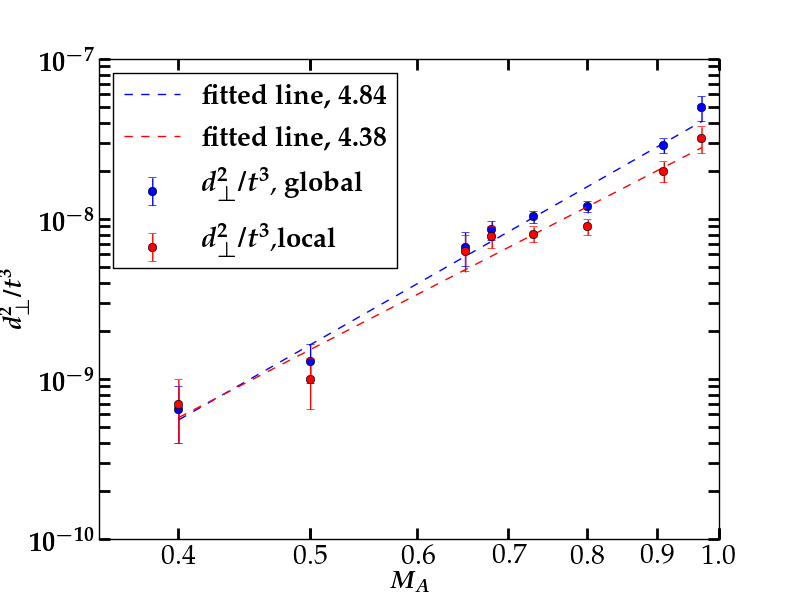}
\end{subfigure}
\begin{subfigure}[]{}
\centering
\includegraphics[width=0.31\textwidth,height=0.2\textheight]{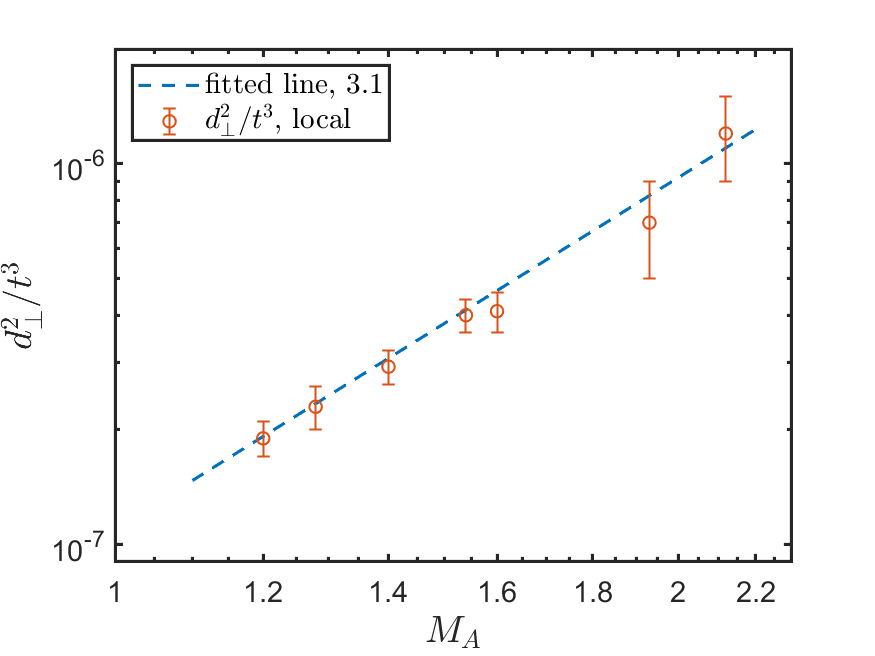}
\end{subfigure}
\caption{{\em a)} The ratio between perpendicular and parallel diffusion ($D_\perp/D_\parallel$) when $\lambda_{\|} < L$. Transport of particles in both the whole turbulence (blue) and Alf\'ven modes (red) are presented. The fitting lines and power law indices are marked in the legend. $M_A$ dependence of superdiffusion of particles in Alfv\'en modes. {\em b)} $M_A\leq 1$. The blue line is the fit in the global reference frame. The red line shows the fit for the data points obtained in the local reference frame. {\em c)} in super-Alfv\'enic turbulence. The calculation is done in local reference frame. From \cite{Maiti2021}.}\label{fig:diffindexvsMA}
\end{figure}

The dependence of the perpendicular diffusion coefficient on Mach number is presented in Fig.~\ref{fig:diffindexvsMA}a for both total turbulence data cubes and the Alfv\'en modes. The relation between the diffusion coefficients and Alfv\'enic Mach numbers is fitted by a power law:$D_{\perp}/D_\| \propto M_A ^\zeta$. The diffusion coefficients are compared for both the Alfv\'en modes and the total turbulence. The fitting index is $3.65$ for the total turbulence data cubes and $3.83$ for Alfv\'en modes. Both results are compatible with the theoretical result from \cite{YL08} based on the anisotropy of Alfv\'en modes. For both regimes where CRs' mean free path is larger and smaller than the injection scale, the results from Alfv\'en modes are closer to the expected index $\zeta=4$ than those from total turbulence data cubes \citep{Maiti2021}. This is due to the contributions from the magnetosonic modes in the total turbulence data cubes. CR perpendicular diffusion is strongly dependent on the Alfv\'enic Mach number, and it is essential to consider the anisotropy of MHD turbulence when modelling CR propagation.

\subsection{Superdiffusion: local vs. global reference frames}
\begin{figure}[h]
\includegraphics[width=\columnwidth]{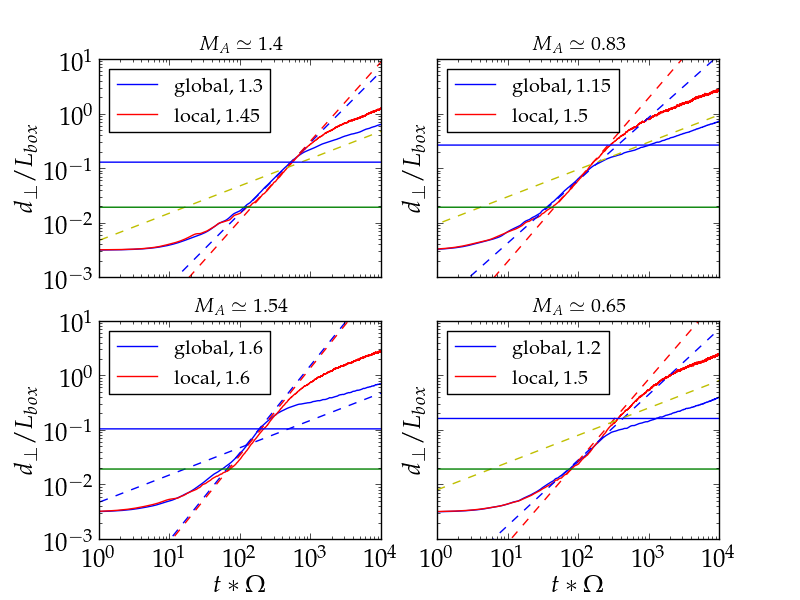}
\caption{ Perpendicular transport of CRs on small scales. The y axis represents the perpendicular distances normalised by the box length ($d_\perp/L_{box}$) and the x axis represents the CR gyro periods ($t*\Omega$). The perpendicular distances obtained from numerical simulations are represented in the global (blue lines) and the local (red lines) reference frame. The horizontal lines in the plots represents the inertial range of turbulence. The yellow lines represents the reference line for normal diffusion with a slope of 0.5. From \cite{Maiti2021}.}\label{fig:dperp_cho}
\centering
\end{figure} 

\begin{figure}[h]
\begin{subfigure}[]{}
\includegraphics[width=0.31\columnwidth,height=0.2\textheight]{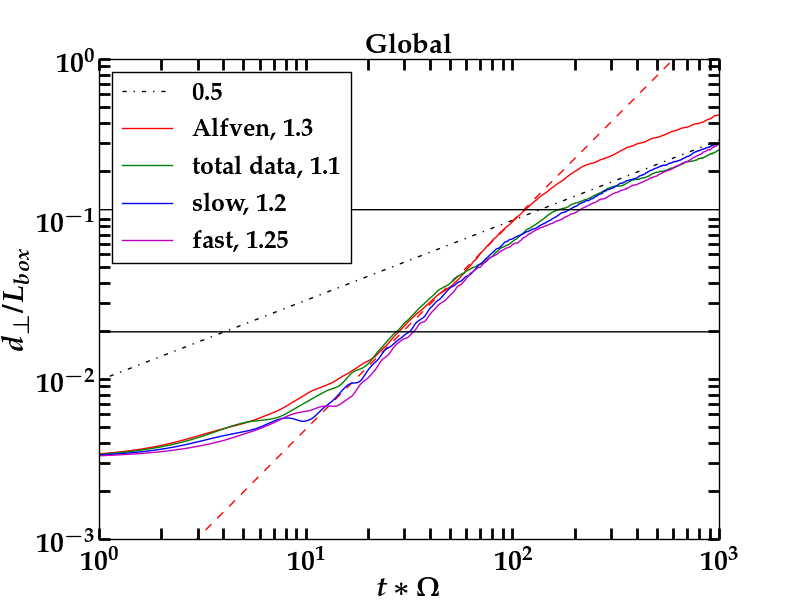}
\end{subfigure}
\begin{subfigure}[]{}
\includegraphics[width=0.31\columnwidth,height=0.2\textheight]{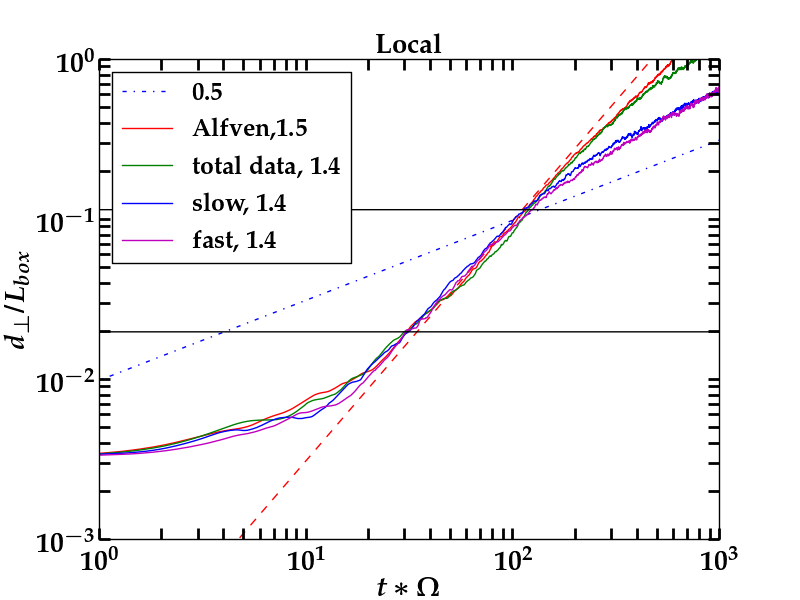}
\end{subfigure}
\begin{subfigure}[]{}
\includegraphics[width=0.33\columnwidth,height=0.2\textheight]{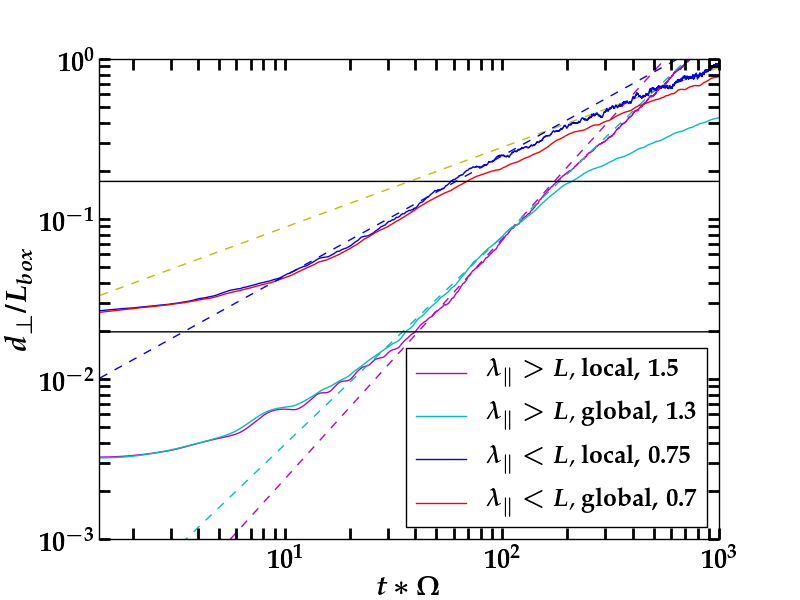}
\end{subfigure}
\centering
\caption{Superdiffusion in different regimes. a,b) Results in both {\em global} and {\em local } magnetic reference frames with decomposed modes included for comparison. c) with $\lambda_{\|} > L$ and with $\lambda_{\|} < L$ in total turbulence data and Alfv\'en modes, respectively. From \cite{Maiti2021}.}
\label{fig:dperp_dt}
\centering
\end{figure}

In this subsection, we discuss the particle transport on small scales within the inertial range. The time evolution for the perpendicular transport can be characterized by a power law: $d_\perp\propto t^{\alpha}$. The Richardson diffusion describes the explosive growth of the separation of particles in turbulence medium, as inferred from fluids experiments many decades ago \citep{Richardson1926}. Richardson law is equivalent to the Kolmogorov spectrum. Therefore Richardson diffusion is also expected in MHD turbulence since the perpendicular spectrum of Alfv\'enic turbulence has a Kolmogorov scaling \citep{GS95}. The Richardson diffusion in MHD turbulence was confirmed with high resolution numerical simulations by \cite{Eyink_NAT}. Following the Richardson diffusion of magnetic field lines, CRs are also expected to undergo superdiffusion on the scales below the injection scale with the index over time $\alpha={3/2}$ \citep{YL08, LY14}.  

The perpendicular distance transport is plotted in Fig.~\ref{fig:dperp_cho} for sub-Alfv\'enic and super-Alfv\'enic total turbulence data. Fig.~\ref{fig:dperp_cho} demonstrates the CR transport in both {\em global} and {\em local} magnetic reference frames. All cases can be fitted with the index close to $\alpha\sim1.5$ in the {\em local} reference frame, in line with the theoretical expectations since the global magnetic field generally differs from local magnetic fields in turbulent medium. On the other hand, $\alpha$ is only close to 1.5 at super-Alfv\'enic cases, decreasing substantially and close to $1$ for sub-Alfv\'enic cases in {\em global} reference frame. 

The perpendicular transport in decomposed MHD modes are demonstrated in Fig.~\ref{fig:dperp_dt}a,b. Turbulence data cubes with Alfv\'enic Mach numbers ranging from $0.4$ to $1.0$ are considered. Superdiffusion is generally observed in all the tests. The particle transport in decomposed Alfv\'enic modes are the closest to the Richardson diffusion (index $\alpha=1.5$) in the {\em local} reference frame compared to the other modes. In global magnetic reference frame, the indices deviate further from the Richardson diffusion as expected. The similar trend is observed in super-Alfv\'enic turbulence as well~\citep{Maiti2021}. Fig.~\ref{fig:dperp_dt}c compares the superdiffusion diffusion of CRs with mean free path $\lambda_{\|}$ larger or smaller than the turbulence injection scale $L$, respectively. The CR transport shows Richardson diffusion in the {\em local} reference frame, i.e., with an power law index of $1.5$ when $\lambda_\|>L$ and reduced to $0.75$ when $\lambda_\|<L$, in line with the theoretical expectation \cite{LY14}. Fig~\ref{fig:diffindexvsMA}b,c demonstrates the dependence of superdiffusion in sub-Alfv\'enic regime. The fitting power law index in the {\em local} reference frame ($4.34$) is closer to the theoretical expectation $M_A ^4$ than that in the global frame ($4.84$). For the super-Alfv\'enic turbulence, the super-diffusion on small scales ${d_{\perp}}^2 / t^3 $ in the {\em local} reference frame show a dependence of $M_A^{3.1}$, close to the $ M_A^3$ theoretical relation \cite{LY14}.

The cross field transport is much determined by Alfv\'en modes, all the tests performed with Alfv\'en modes show better consistency with theoretical predictions earlier \citep{YL08, LY14}. Particularly, the results obtained in the {\em local} reference frame are completely in line with the theory. Then depending on the degree of Alfv\'enicity, the observed cross field transport property can vary. For instance, the super-diffusion index can be different from the Richardson diffusion one.
The actual observed superdiffusion index can vary depending on the modes composition and Alfv\'enic Mach number of local turbulence, which explains the observed variety of superdiffusion indices in both solar wind \citep[see, e.g.][]{PerriZimbardo2009} and the supernova remnants \citep{Perri2016}. 

\subsection{The role of small scale instabilities} 

In addition to the large scale turbulence, small scale instability generated perturbations also play crucial roles. Particularly at shock front, studies of instabilities have been one of the major efforts in the field since the acceleration efficiency is essentially determined by the confinement at the shock front and magnetic field amplifications. In fact, the small scale instabilities and large scale turbulence are not independent of each other. First of all, the instability generated waves can be damped through the interaction with the large scale turbulence \cite{YL02, FG04}. Secondly, the large scale turbulence also generates small scale waves through firehose, gyroresonance instability, etc. \cite[see][]{Schekochihin2005, LB06, YL11, Lebiga18}.

Instabilities driven by CR momentum anisotropy or inhomogeneity can amplify the local mean magnetic field, increasing the scattering rate of CRs and regulating the CR transport. These effects are of key importance in the context of CR acceleration in SNR shocks, propagation around acceleration sources and in the Galaxy. The most important instabilities frequently considered are incompressible slab modes driven by the anisotropy in the CRs momentum distribution, and a unified general linear dispersion relation for these modes was derived \cite{Bykov2013}. However, numerical simulations are necessary to assess the non-linear evolution and saturation of these instabilities. In the vicinity of acceleration sites (shocks, magnetic reconnection sites), the confinement through the streaming instability is more effective because of much enhanced streaming flux. A natural way to increase the scattering rate is through both the resonant and non-resonant streaming instabilities~\cite[see][]{Wentzel74,Cesarsky80,Bell2004,Malkov2013}. 
 
The diffusive propagation of CRs in the Galaxy, far from sources, is considered to be regulated by the interactions with the interstellar medium (ISM) background turbulence, as discussed in previous subsections. However, CRs with energies below ~100 GeV are mainly influenced by the self-generated instabilities because of enhanced flux in the low energy range. The compression/shearing by the large scale ISM turbulence can produce deformations in the local particle pitch angle distribution due to the conservation of the first adiabatic invariant. Such anisotropic distribution is subjected to various instabilities.  While the hydrodynamic instability requires certain threshold, the kinetic instability can grow faster with small deviations from isotropy. This anisotropy in the momentum distribution induces gyroresonance instability. Unlike the streaming instability, the gyroresonance instability does not require the bulk motion of CRs. The wave grows at the expense of the free energy from CRs, provided by the large scale turbulence motions. In the case that the energy growth rate reaches the turbulence energy cascading rate, turbulence is damped. This is one of the feedbacks from CRs on turbulence~\cite{LB06}. An analytical equilibrium model of the CR diffusion generated by this instability is proposed by \cite{YL11} based on the QLT, and the dependence of the diffusion coefficients is derived with the parameters of the ISM turbulence. They found this mechanism to be important for the propagation of CRs (<100GeVs) in collisionless medium, such as the Halo and Hot Ionized Medium of Galaxy, and ICM of galaxies. 

\begin{figure}[h]
\centering
\begin{subfigure}[]{}
         \centering
         \includegraphics[width=0.54\textwidth,height=0.3\textheight]{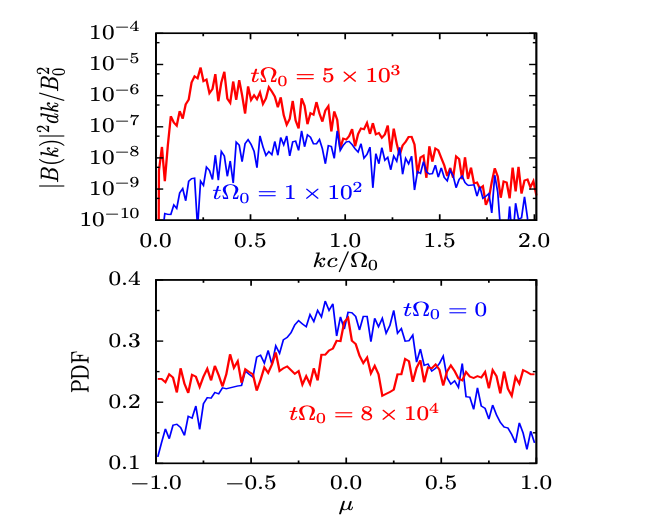}
              \end{subfigure}
            \hfill
\begin{subfigure}[]{}
         \centering
         \includegraphics[width=0.42\textwidth,height=0.28\textheight]{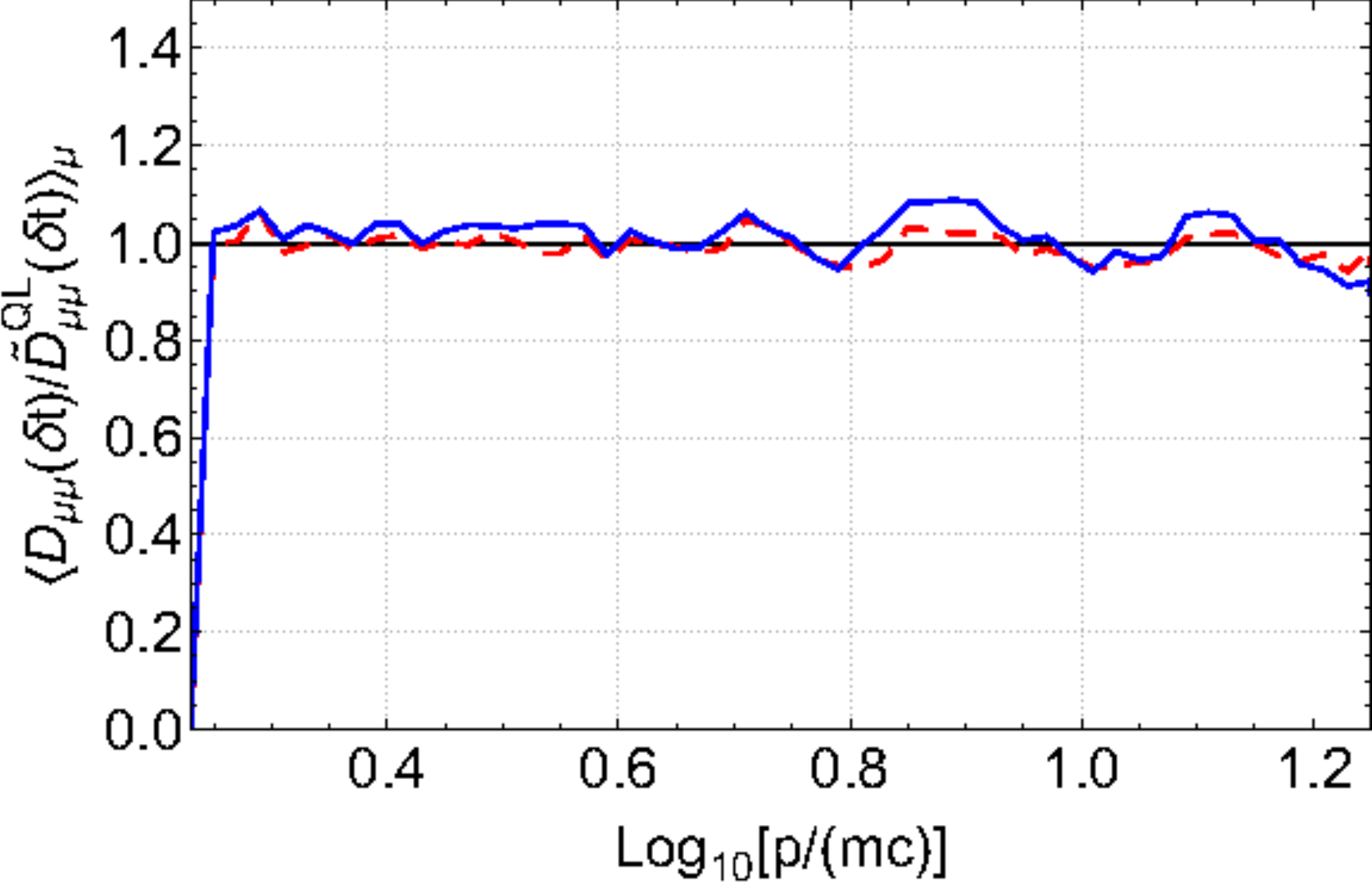}
              \end{subfigure}
     \caption{{\em a)} Normalized power spectrum of magnetic field $|B(k)|^2/B_0^2$ ({\it left column}) and PDF of particle pitch angle cosine; b) $\langle D_{\mu\mu}(\delta t) / \widetilde{D}_{\mu\mu}^{QL} (\delta t) \rangle_{\mu}$ ratio ({\it right column}) for models
starting with anisotropy $A_0$: $+0.1$. Lines with different colors indicate different times: 
$t \Omega_0 = 10^3$ ({\color{red} red {\color{magenta} dashed} line}), 
$t \Omega_0 = 2 \times 10^3$ ({\color{blue} blue {\color{magenta} solid} line}).
In the power spectrum plots, the region where the numerical dissipation dominates 
are gray-shaded (wavelenghts $\le 32$ grid cells); the blue shaded region represents the interval $\Omega_0 m_{cr} / p_{\max} < k < \Omega_0 m_{cr} / p_{\min}$. From \cite{Lebiga18}.}
\label{fig:long_spectrum}
\end{figure}

Numerical study was conducted on the role of gyroresonance instability on CR scattering \cite{Lebiga18}. 
The pitch-angle diffusion coefficient averaged during the growth phase of the instability is found to be in good agreement with the QLT estimates for static waves (Fig.\ref{fig:long_spectrum}). This result laid a solid foundation for further investigations on the role of the CR gyroresonance instability, which together with streaming instability dominate the transport of low energy CRs.

The development of the above highlighted CR instabilities is in general restricted by different wave damping mechanisms. Inside molecular clouds, where the gas ionization rate is low, the ion-neutral damping can suppress the resonant waves, then increasing the CRs diffusivity in the interior of the cloud \cite{KulsrudPearce1969,Everett2011}. In more ionized medium (for example, close to the acceleration sources and in the galactic Halo), the damping mechanism is probably provided by the nonlinear Landau damping \cite{Kulsrud1978} or by the large scale MHD turbulence cascade \cite{FG04, YL04}. Undoubtedly, the self-regulated escape from source and propagation in the galaxy accounting for the wave damping requires dedicated studies with the nonlinear theory and numerical simulations.

\subsection{Local particle transport near sources}

While global observables from direct measurement of cosmic rays offer insight on general properties of CRs, such as the age, abundance, etc., they do not inform us much about the microphysics of plasma interactions between particles and turbulence. This is because the cosmic rays measured on Earth have been going through all the interstellar volume from their sources, which are nontrivial to pinpoint because of the frequent scattering and diffusion process that CRs have experienced. The transport properties we obtain from fitting the global observables with the propagation codes therefore only give us an averaged quantities smearing out much of the variation of turbulence properties in interstellar medium and the corresponding physics of turbulence particle interaction.      

To understand the microphysics, local particle transport and associated phenomena are the ones that matter. For supernova remnants, gamma ray emmission from molecular clouds (MCs) near SNRs provide one of the best ways to obtain information on the accelerated protons. As a matter of fact, the $\gamma$-ray flux coming from MCs associated to SNRs were observed to be 1-2 orders of magnitude more intense compared to isolated MCs in the ISM, which are only targeted by the background of CRs~\citep[e.g.][]{Fujita2009,Yan2012}. Recently, high precision measurements of pulsar wind nebulae have also opened up a new window to the study of particle transport in local environments of PWNs \citep[see, e.g., ][]{LYZ19, LY20}. In particular, observation of the High-Altitude Water Cherenkov Observatory (HAWC) has revealed a TeV gamma-ray halo around the Geminga pulsar, with a spatial extension of $\gtrsim 30$\,pc \citep{HAWC17_Geminga}. The TeV emission is believed to arise from cosmic-ray electrons/positrons injected from the pulsar wind nebula (PWN), via inverse-Compton (IC) scattering off cosmic microwave background (CMB) photons. The detection of such a diffuse TeV emission has been interpreted as the presence of a slow diffusion zone around the pulsar \citep{HAWC17_Geminga}, which has spurred immense interest in the community. Different proposals have been put forward \cite{Profumo18,Xi18,Recchia2021}, which shows the urgent need of understanding of basic plasma processes in the astroparticle physics. It restates that the picture of simple uniform diffusion dictated by the same hydrodynamic Kolmogorov turbulence is far from reality. As we discussed extensively above, turbulence is magnetohydrodynamic and varies substantially from place to place depending on local environments, particularly forcing mechanisms. 

Multi-wavelength study holds the best chance to understand the local transport and the associated physical processes. As discussed above, multiwavelength analysis augmented by advanced data analysis based on understanding of underlying plasma properties points to the origin of the gamma ray enhancement in the center of the Cygnus-X region. For Geminga, an upper limit of X-ray flux has been obtained from observations by XMM-Newton and Chandra around the pulsar. Since X-ray is supposed to arise from synchrotron radiation of the same electrons that account for the TeV emission, the X-ray upper limit translates to an upper limit for the magnetic field strength in the TeV halo, i.e., $\leq 0.8\mu$G, which is significantly weaker than the typical interstellar medium (ISM) magnetic field. The combination of a small diffusion coefficient and a weak magnetic field makes the two zone scenario unfavorable. 

\begin{figure}[h]
\centering 
\begin{subfigure}[]{}
         \centering
         \includegraphics[width=0.48\textwidth]{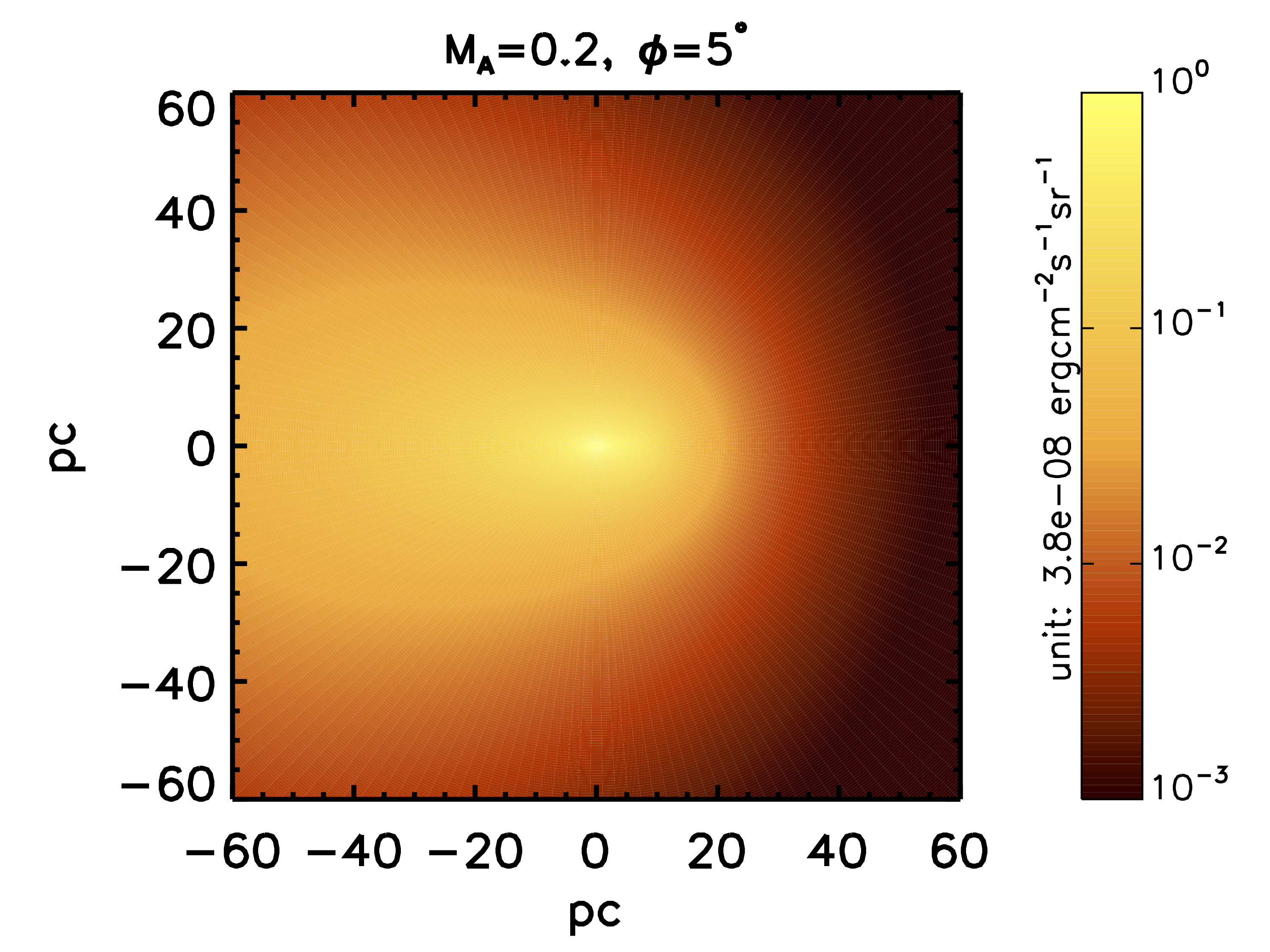}
              \end{subfigure}
     \hfill
     \begin{subfigure}[]{}
         \centering
         \includegraphics[width=0.48\textwidth,height=5.4cm]{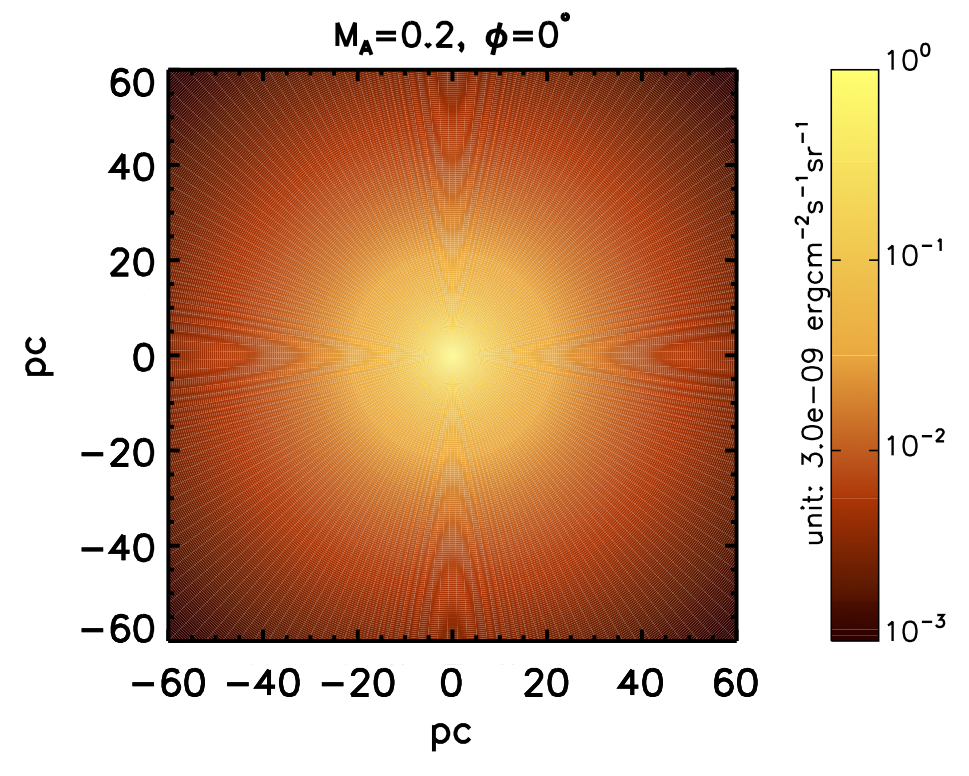}
              \end{subfigure}
     \hfill
     \\
\begin{subfigure}[]{}
         \centering
         \includegraphics[width=0.48\textwidth]{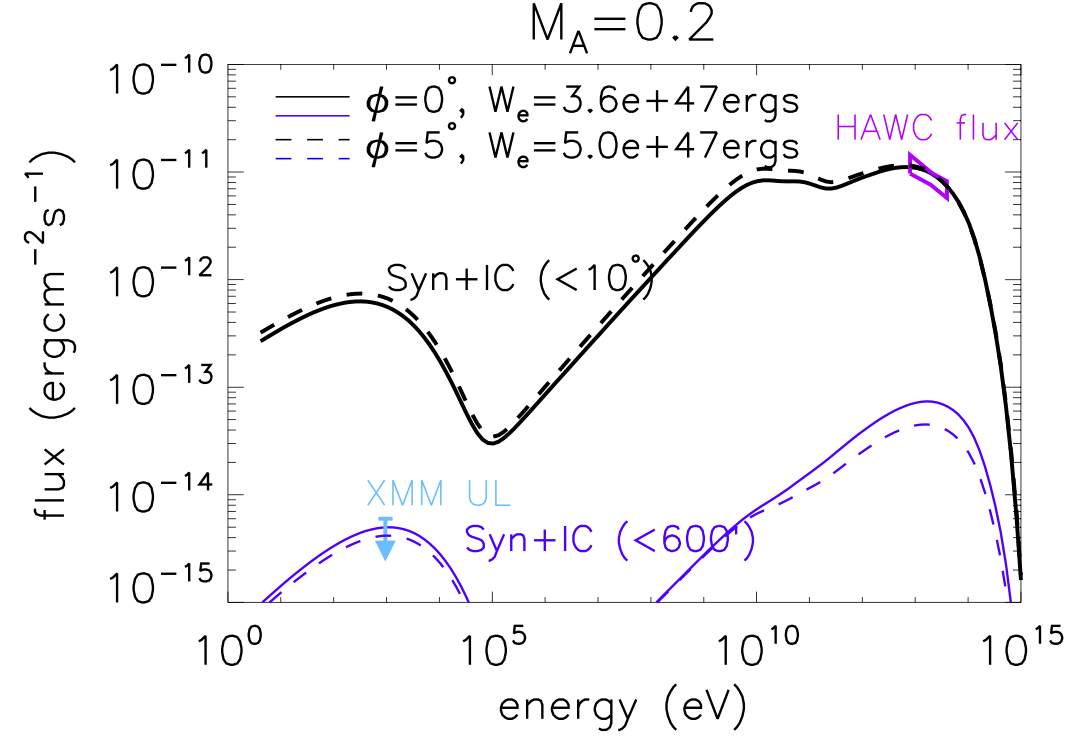}
              \end{subfigure}
              \hfil
\begin{subfigure}[]{}
         \centering
         \includegraphics[width=0.48\textwidth]{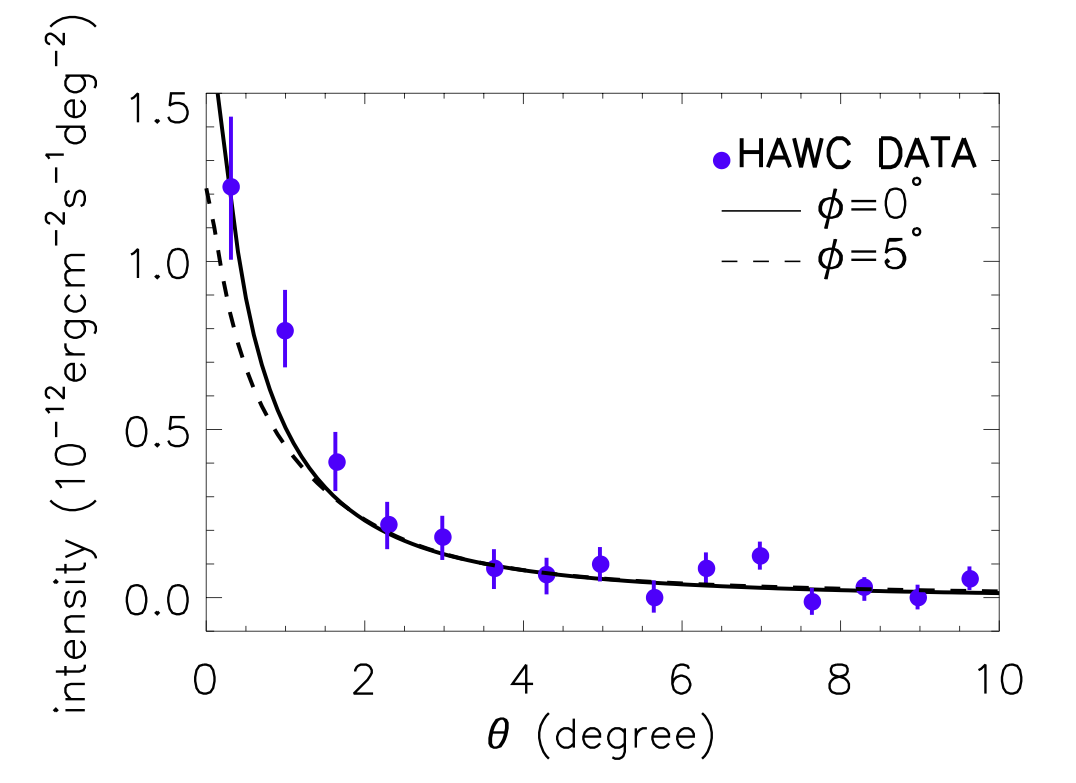}
              \end{subfigure}
\caption{{\em a,b)} Predicted $8-40$\,TeV SBP with different Alfv{\'e}nic Mach number $M_A=0.2$ and different viewing angle $\phi=0^\circ,5^\circ$. {\em c,d)} Results with $\phi=0^\circ$ and $\phi=5^\circ$ for $M_A=0.2$. {\em c)} the predicted multiwavelength flux from a region within $10^\circ$ from Geminga (black curves) and from a region within $600''$ from Geminga (blue curves). Solid curves represent the result of $\phi=0^\circ$ while dashed curves represent the result of $\phi=5^\circ$. The magenta bowtie and the cyan arrow represent the flux measured by HAWC and the upper limit from XMM-Newton respectively. {\em d)} the predicted 1D ($\zeta$-averaged) SBP in $8-40$\,TeV in comparison with the measured one by HAWC, which is shown as blue circles. From \cite{LYZ19}.}
\label{Geminga}
\end{figure}

On the other hand, the magnetic field in ISM generally has a mean direction within one coherent length, which is typically $\sim 50-100\,$pc \citep{Chepurnov10, Beck16} and comparable to the size of the TeV halo around Geminga.  1D isotropic particle diffusion does not necessarily hold in this scenario, since particles diffuse faster along the mean magnetic field than perpendicular to the mean magnetic field if it exists. In the case of $M_A<1$, the perpendicular diffusion coefficient is given by $D_{\perp}=D_{\|}M_A^4$ as shown in \S\ref{largescale_transport} (Fig.\ref{fig:diffindexvsMA}a). Also, the synchrotron radiation intensity becomes anisotropic. Electrons that move along the magnetic field will radiate much less efficiently than those move perpendicular to the magnetic field. Therefore, if the mean magnetic field in the vicinity of Geminga has small inclination toward our line of sight (LOS), the observed synchrotron radiation flux would be much reduced compared to that with the assumption of an isotropic magnetic field, while the diffusion perpendicular to the LOS is slow as suggested by the TeV observation. It is demonstrated in {\cite{LYZ19}} that both X-ray and TeV observations can be both fit with typical conditions for ISM, such as the magnetic field, the diffusion coefficient and the field perturbation level, by considering anisotropic particle diffusion which is a natural outcome in the presence of sub-Alfv{\'e}nic turbulence (Fig.\ref{Geminga}). The viewing angle plays an important role in determining the observation signals. The work shows the importance of accounting for the 3D structure of the local MHD turbulence, particularly their anisotropy in treating particle transport. 

\section{Summary}
We have reviewed some of the recent developments on the understanding of MHD turbulence and its interaction with CRs. Theoretical developement alongside with observational discoveries are presented. We emphasize that CR research is synergetic with study of turbulence. Our major conclusions are:
\begin{itemize}
    \item Galactic turbulence has 3D structure and profile. 1D approximation is inadequate, particularly in relation to CR physics.
    \item The composition of MHD turbulence depends on the driving mechanism. Alfven modes dominate in the case of solenoid driving, whereas MS modes prevail with compressible driving.
    \item Both Alfven and MS modes are detected in star-forming area and SNR.
    \item Compressible fast modes do not have a weak regime and are generally isotropic. They dominate CR transport through direct scattering, therefore. Near sources, and for GCRs < a few hundred GeV, plasma instabilities are more important.
    \item Multi-waveband study holds the key to CR research. In Cygnus X, the $\gamma$-ray cocoon largely coincides with the Compressible modes dominant zone, consistent with the theoretical prediction on the dominant role of MS modes on particle transport.
    \item The efficiency and energy dependence of CR scattering depend on local turbulence properties dictated by turbulence driving and damping/medium parameters. CR transport is thus inhomogeneous.
    \item CR perpendicular transport is diffusive in large scale turbulence (w. $D_\perp /D_{||}  \propto M_A^4$) and superdiffusive on small scales. The actual superdiffusion index varies with the modes composition and Alfv\'enic Mach number of turbulence.
\end{itemize}

{\small
\bibliography{yanCR.bib}
\bibliographystyle{JHEP}
}

%
%
%

\end{document}